\documentclass[reprint,
superscriptaddress,
 twocolumn,
 nofootinbib,
 nobibnotes,
 amsmath,amssymb,
 aps,
prl,
 floatfix,
 tightenlines,
 10pt
]{revtex4-1}

\bibpunct{[}{]}{;}{n}{}{}
\usepackage{graphicx}
\usepackage{dcolumn}
\usepackage{bm}
\usepackage{physics}
\usepackage{float}
\usepackage{caption}
\usepackage{subcaption}
\usepackage{soul}
\usepackage{xcolor}
\usepackage{bbold}
\usepackage{nicefrac}
\usepackage{xcolor}
\usepackage{amsmath,amsthm, amsfonts,amssymb}
\usepackage{mathtools}%

\usepackage[font=small,labelfont=bf]{caption}

\RequirePackage[colorlinks=true,allcolors=blue,
unicode=true,hypertexnames=false]{hyperref}
\hypersetup{pdfstartview={XYZ null null 1.25}}

\begin{document}
\title{
Magnetic signatures of domain walls in $s+is$ and $s+id$ superconductors: observability and what that can tell us about the superconducting order parameter}

\author{Andrea Benfenati} 
\affiliation{Department of Theoretical Physics, The Royal Institute of Technology, Stockholm SE-10691, Sweden}
\author{Mats Barkman} 
\affiliation{Department of Theoretical Physics, The Royal Institute of Technology, Stockholm SE-10691, Sweden}
\author{Thomas Winyard} 
\affiliation{School of Mathematics, University of Leeds, Leeds LS2 9JT, United Kingdom}
\author{Alex Wormald}
\affiliation{School of Mathematics, University of Leeds, Leeds LS2 9JT, United Kingdom}
\author{Martin Speight}
\affiliation{School of Mathematics, University of Leeds, Leeds LS2 9JT, United Kingdom}
\author {Egor Babaev}
\affiliation{Department of Theoretical Physics, The Royal Institute of Technology, Stockholm SE-10691, Sweden}

\begin{abstract}
One of the defining features of spontaneously broken time-reversal symmetry (BTRS) is the existence of domain walls, the detection of which would be strong evidence for such systems. There is keen interest in BTRS currently, in part, due to recent muon spin rotation experiments, which have pointed towards $\textrm{Ba}_{1-x}\textrm{K}_x\textrm{Fe}_2\textrm{As}_2$ exhibiting a remarkable case of $s$-wave superconductivity with spontaneously broken time-reversal symmetry. A key question, however, is how to differentiate between the different theoretical models which describe such a state. Two particularly popular choices of model are $s+is$ and $s+id$  superconducting states. In this paper, we obtain solutions for domain walls in $s+is$ and $s+id$ systems, including the effects of lattice anisotropies. We show that, in general, both models exhibit spontaneous magnetic field, that extend along the entire length of the domain wall. We demonstrate the qualitative difference between the magnetic signatures of $s+is$ and $s+id$ domain walls and propose a procedure to extract the superconducting pairing symmetry from the magnetic-field response of domain walls.\end{abstract}

\date{August 21, 2019}
\maketitle


\section{Introduction}
Superconducting states that spontaneously break time reversal symmetry have been a subject of experimental pursuit and theoretical investigation over the past few decades. Although a number of  candidate materials  was discovered, the nature of their order parameters remains a subject of debate.
Recent experimental works in iron-based superconductors reported spontaneous breakdown of time- reversal symmetry (BTRS) in $ \textrm{Ba}_{1-x}\textrm{K}_x\textrm{Fe}_2\textrm{As}_2$ \cite{Grinenko2017, grinenko2018emerging}, based on muon spin rotation measurements. 
The leading candidates for the BTRS state in Fe-based compounds are $s+is$ and $s+id$ states \cite{StanevTesanovic,Lee.Zhang.Wu:09,Carlstroem2011a,Maiti2013,Boeker2017,platt2012}.  The experiments \cite{Grinenko2017, grinenko2018emerging}, detected spontaneous magnetic fields appearing in superconducting states. These are believed to be a hallmark of the spontaneous breakdown of time reversal symmetry.

It was suggested that impurities generate a magnetic field in $ s+id $ superconductors \cite{Lee.Zhang.Wu:09}.
It has also been pointed out that, in contrast, isotropic $s+is$ superconductors exhibit no such effect for a spherically symmetric impurity
as well as no magnetic signatures of straight domain walls. Spontaneous magnetic fields appear in an isotropic system if one creates cross-gradients of
relative density and relative phase \cite{Garaud2014}. Such configurations arise when domain walls interact
with pinning centers or the boundary of the sample \cite{Garaud2014}.
Several proposals were made to distinguish between $s+is$ and $s+id$ states from various
configurations of impurities
\cite{ChubukovMaitiSigrist,lin2016distinguishing,
Silaev.Garaud.ea:15,Garaud2016,garaud2018properties}.
In the most recent proposal \cite{vadimov2018polarization} it was suggested that for models relevant for $ \textrm{Ba}_{1-x}\textrm{K}_x\textrm{Fe}_2\textrm{As}_2$  one can distinguish
between $s+is$ and $s+id$ superconductors. It also supported the material being an $s+is$ state \cite{grinenko2018emerging}.

However, determining the nature of BTRS states is an extremely difficult task. Proposed experimental signatures are 
vortex clustering, flux flow viscosity at the BTRS phase transition
\cite{garaud2018properties,Silaev.Babaev:13}, soft collective modes close to the transition \cite{Lin2012,Carlstroem2011a, marciani2013legett,Maiti2013}, formation of metastable Skyrmions \cite{Garaud2011,Garaud2013,Garaud2014},
and quasi-particle interference \cite{Hirschfeld2015,Boeker2017}.

In this paper, we focus on a simple feature to measure and compare, showing how the states can be diagnosed via
the observation of the magnetic field of domain walls separating $s+is$ and $s-is$ or $s+id$ and $s-id$ domains.
This can be observed in superconducting quantum interference device (SQUID), scanning Hall probes \cite{moler1,moler2,moshchalkov2,gutierrez2012scanning}, and muon spin rotation \cite{grinenko2018emerging}.

Although, in the isotropic $s+is$ models, a straight domain wall does not 
produce any magnetic field \cite{Garaud2014}, it was observed 
in Refs. \cite{silaev2017non,winyard2018,winyard2018skyrmion} that in the presence of anisotropies, the phase difference between the components couples directly to the magnetic
field, which could lead to domain walls exhibiting spontaneous magnetic fields in $s+is$ states of the anisotropic materials such as $ \textrm{Ba}_{1-x}\textrm{K}_x\textrm{Fe}_2\textrm{As}_2$.

In this paper, we study the spontaneous magnetic field generated by pinned domain walls, as a function of their orientation with respect to the crystalline axes.
We demonstrate that if the domain wall exists in one of the crystalline planes in either $s+is$ or $s+id$ systems, then there is only a localized effect on the boundary, caused by the pinning sites geometry. However, if the domain wall is not aligned with the crystalline planes then a much  stronger spontaneous magnetic-field signature can be observed. Importantly, this magnetic response, unlike that caused by the pinning geometry, extends along the entire length of the domain wall and is dependent on the direction of the domain-wall normal. However, there is a very different directional dependence for $s+is$ and $s+id$ superconductors, which is connected directly to the underlying pairing symmetry of the system. For simplicity, we will refer to such a spatially extended magnetic response as the {\it bulk} magnetic field. It is proposed that this behavior can be studied experimentally to determine the pairing symmetry of the superconducting state.

\section{Ginzburg-Landau Formulation}
We are interested in a clean three-band microscopic model which has been proposed to describe the BTRS superconducting state in iron-based superconductors (such as $ \textrm{Ba}_{1-x}\textrm{K}_x\textrm{Fe}_2\textrm{As}_2$) \cite{Maiti2013, Marciani2013legett
, StanevTesanovic} with three coupled microscopic order parameters $\Delta_1,\Delta_2,\Delta_3$. The associated microscopic coupling matrix $\hat{\Lambda}$ yields time reversal symmetry breaking and is dominated by competing interband repulsive pairing
\begin{equation}\label{eq:interactionMatrix}
    \hat{\Lambda} =- \mqty(0 & \eta & \lambda \\ \eta & 0 & \lambda \\ \lambda & \lambda & 0),
\end{equation}
where $\lambda$ and $\eta$ are positive definite. This interaction matrix can describe both $s+is$ and $s+id$ superconductors. To study these states, we use a microscopically derived multiband Ginzburg-Landau (GL) free-energy functional \cite{Garaud2016,garaud2017microscopically}. The derivation is summarized in the Appendix. Multiband superconductors are described in the Ginzburg-Landau framework by complex classical fields $\psi_\alpha=\abs{\psi_\alpha}e^{i\theta_\alpha}$, however the number of Ginzburg-Landau fields does not always coincide with the number of microscopic gaps. This generally depends on the specific choice of the coupling matrix $\hat{\Lambda}$ as described in Refs. \cite{Maiti2013,Marciani2013legett} and in the Appendix. In our case, since  $\hat{\Lambda}$ describes a system dominated by repulsive interband pairing, the Ginzburg-Landau free energy for $s+is$ and $s+id$ superconductors is an effective two-component model,
\begin{align} \label{eq:FreeEnergyDensity}
	F &= \int \dd^3 x \qty{
	\qty(\Pi_i\psi_\alpha)^*
	Q^{\alpha\beta}_{ij}\qty(\Pi_j\psi_\beta) + \mathcal{V}_\textrm{p}
     + \frac{\qty(\curl{\vb{A}})^2}{8\pi}},
\end{align}
where we imply summation over repeated indices. The Latin indices $i,j \in \qty{x,y,z}$ label spatial components and Greek indices $\alpha,\beta\in \qty{1,2}$ label the Ginzburg Landau order parameters, described by the complex fields $\psi_\alpha\qty(x,y,z)=\abs{\psi_\alpha}e^{i\theta_\alpha}$. The covariant derivative is given as $\Pi_j= \partial_j + iqA_j$ where $\vb{A}$ is the magnetic vector potential.

The spatial symmetry of the system determines how the anisotropy tensors $Q^{\alpha\beta}_{ij}$ in Eq. \eqref{eq:FreeEnergyDensity} couple the covariant derivatives acting on the matter fields. For the free energy to be real, we have $Q^{\alpha\beta}_{ij}=Q^{\alpha\beta}_{ji}$ and $Q^{\alpha\beta}_{ij}=Q^{\beta\alpha}_{ij}$. An additional constraint on the sign and magnitudes of the tensor components guarantees $F$ being bounded below.
Finally, $\mathcal{V}_\textrm{p}$ in Eq. \eqref{eq:FreeEnergyDensity} is the potential density. This term determines the possible ground states and is responsible for BTRS. The potential density reads
\begin{align}\label{eq:potential}
\begin{split}
    \mathcal{V}_\textrm{p} &= \sum_\alpha^2 a_i\abs{\psi_\alpha}^2 + \frac{b_\alpha}{2}\abs{\psi_\alpha}^4\\
	 &+\gamma\abs{\psi_1}^2 \abs{\psi_2}^2 +  \frac{\delta}{2}\qty(\psi_1^{*2}\psi_2^{2} + \psi_1^{2}\psi_2^{*2}),
\end{split}
\end{align}
where $\delta > 0 $ and $a_\alpha = a_\alpha(T/T_c)$ is temperature dependent, where $T_c$ is the critical temperature. The coefficients used in Eqs.\eqref{eq:FreeEnergyDensity} and \eqref{eq:potential} are systematically obtained from the microscopic coupling matrix $\hat{\Lambda}$ and from $T/T_c$ as reported in the Appendix.

We are interested in analyzing $s+is$ and $s+id$ states, a key feature of which is a $Z_2$ degeneracy in the ground state. Namely the phase difference between the two condensates in the ground state can take one of two values $\theta_{12}=\theta_1-\theta_2 = \pm \pi/2$, leading to the ground states $(|\psi_1|,|\psi_2|, \theta_{12}) = (u_1,u_2, \pm \pi/2)$ where $u_\alpha$ is a positive constant. By choosing one of these two possible values, the system spontaneously breaks time-reversal symmetry.
 
It is this degeneracy in the ground states that leads to the possibility of domain-wall defects, which are one-dimensional structures interpolating between two ground-state values. This splits the system into two domains, each in a different ground state with the domain wall as an interface between them. As our theory is formed of continuous fields, the phase difference must interpolate smoothly from $\theta_{12}=\pi/2$ to $\theta_{12} = - \pi/2$. In isotropic superconductors, domain walls are associated with zero magnetic field, unless the domain wall is attached to an inhomogeneous pinning center or there is an underlying density inhomogeneity \cite{Garaud2014,Garaud2016,Silaev.Garaud.ea:15}. However, in the presence of anisotropies, it has been shown that the magnetic field is coupled with phase difference gradients \cite{silaev2017non,winyard2018,winyard2018skyrmion,vadimov2018polarization} and with matter field density gradients \cite{speight2019chiral}.
This would suggest that anisotropies could principally alter the magnetic signatures of domain walls in $s+is$ and $s+id$ systems. Since the experiments \cite{grinenko2018emerging,Grinenko2017} report a $s+is$ state in anisotropic materials \cite{vadimov2018polarization},
this calls for the investigation of domain-wall solutions in anisotropic systems.

The anisotropy tensors for both $s+is$ and $s+id$ superconductors fulfill particular symmetry requirements, stemming from the symmetries of the underlying microscopic theory (this dependence is discussed in the Appendix). In the crystalline axes, both systems consist of all tensors being spatially diagonal ($Q^{\alpha\beta}_{ij} = 0 $ if $i \neq j$) and have $Q^{11}_{xx} = Q^{11}_{yy}$, $Q^{22}_{xx}=Q^{22}_{yy}$. To have a $s+is$ state, it is then necessary that $Q^{12}_{xx}= Q^{12}_{yy}$. If we consider the action of a general rotation acting on $Q^{\alpha\beta}_{ij}$ , it can be shown that the spatial symmetries are $SO(2)\times C_2$, namely, that it has an $SO(2)$ symmetry on the $xy$ plane and a $C_2$ symmetry in the orthogonal direction ($z$ axis). 
The $s+id$ states, on the other hand, requires $Q^{12}_{xx} = - Q^{12}_{yy}$, which leads to the basal $xy$ plane having $C_2$ symmetry and, thus, the three-dimensional system having a $C_2\times C_2$ symmetry. 

Even though the symmetry requirements and the microscopic derivation of the model reduce the number of degrees of freedom, the parameter choice in our model is large. However, we are mostly interested in the qualitative features of the domain walls that would be visible in experiments, as well as ensuring that these features are not fine-tuned. To this end, multiple parameter sets have been considered in tandem with the values shown in this paper. 
We highlight that the value of the $Q^{\alpha\beta}_{ij}$ matrices have no impact on the presence of time-reversal symmetry which only depends on the potential terms of Eq. \eqref{eq:FreeEnergyDensity}.
Finally, for the sake of notation, we introduce the matrix abbreviation $\hat{Q}^{\alpha\beta}$ for the anisotropy tensors, where
\begin{equation}
    \qty(\hat{Q}^{\alpha\beta})_{ij} = Q^{\alpha\beta}_{ij} \,.
\end{equation}
\section{System Setup}
Consider a general domain wall described by the free energy in Eq. \ref{eq:FreeEnergyDensity}. In general, the anisotropy yields a configuration which is dependent on the orientation of the domain wall. This leads to certain orientations being the most energetically favorable and, hence, critical points of the energy functional. 
However, in real materials, due to impurities, spontaneous pinning occurs, so upon cooling a superconductor through the BTRS transition, there will, in general, be pinned domain walls with different orientations.

We propose an experimental setup where pinning centers are introduced on purpose using well-developed experimental techniques, such as
irradiating a sample at a given angle relative to crystalline axes \cite{PhysRevB.75.100504},
or creating dents on its surface that would provide
geometric pinning.
The results of these different procedures are equivalent for our purposes as the ions used for sample irradiation lead to traces in the sample where the superconducting state collapses, which is indistinguishable to a dent within the Ginzburg-Landau formulation. This allows the domain-wall orientation to be fixed experimentally (note that, when we refer to domain-wall orientation, we are always talking about the orientation relative to the crystal axes).
It is important to underline that pinning centers are not necessary for the \textit{bulk} magnetic field to arise, however, they offer a way of controlling the orientation of the domain wall.

\begin{figure}[t]
	\center
	\includegraphics[width=0.80\linewidth]{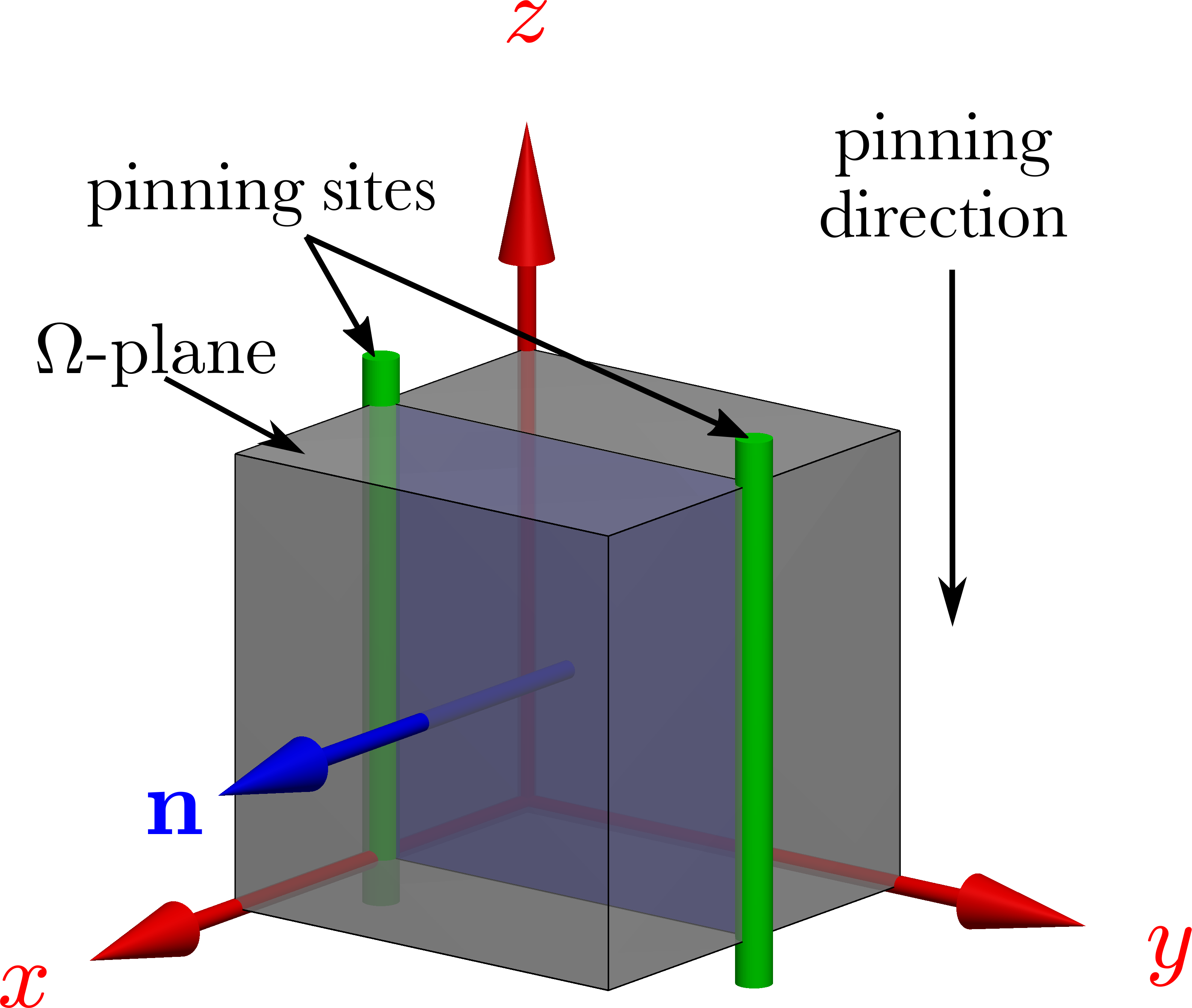}
	\caption{A superconducting sample with a pinned domain wall. The coordinate system labeled by $(x,y,z)$ is the crystalline axes frame. The non superconducting pinning sites are represented by green cylinders. The domain wall, displayed by the blue plane, is then pinned in place by the pinning centers. The direction along which the phase difference interpolates is identified by the domain-wall normal vector $\vb{n}$, displayed by the blue arrow. In this specific case, the domain-wall normal is aligned with the $x$-crystalline axes.}
	\label{fig:setup}
\end{figure}

Below, we focus specifically on the case of a sample that has two columnar  pinning sites, where the superconductivity is suppressed as shown in Fig. \ref{fig:setup} by the green cylinders. If a sample is quenched through 
a phase transition, according to the Kibble-Zurek mechanism \cite{kibble,zurek}, domain walls will form (see also the discussion in Ref. \cite{Garaud2014}). 
A quench-induced domain wall is then captured between these pinning sites as shown in blue, whose orientation is represented uniquely by its normal vector $\vb{n}$. 
As the bulk domain wall is a one-dimensional soliton, the fields only vary in one direction, namely, along the normal $\vb{n}$. This means that, on the blue plane, the domain-wall configuration is translationally invariant. Due to this symmetry, for any chosen parameter set, there is a unique domain-wall solution for any given normal direction $\vb{n}$. Consequently, the normal vector parametrizes the complete family of domain walls for a given system.

If you wish to experimentally consider the domain wall represented by a given $\vb{n}$, you must create two parallel pinning sites where the pinning direction is a vector taken from the plane of the domain wall (namely, it is orthogonal to $\vb{n}$). This then determines a unique direction that is orthogonal to both $\vb{n}$ and the pinning direction, which gives a vector that will lie between the pinning sites (namely, point from one pinning site to the other).
For example, in Fig. \ref{fig:setup}, the normal of the domain wall is in the $x$ direction, hence, the domain wall exists on the $yz$ plane. We can select any direction on the $yz$ plane for the pinning direction, say $z$,  which then picks out that the pinning sites must be separated along the $y$ axis.

To study the bulk domain wall, far from the pinning sites, it would be sufficient to dimensionally reduce the system to a one-dimensional system due to translation invariance orthogonal to $\vb{n}$. However, it is useful to consider the interaction of domain walls with the pinning sites. Hence, if we include the pinning sites, the system has translation invariance only along the pinning direction. Therefore, it is sufficient to consider a two-dimensional (2D) domain $\Omega$, orthogonal to the pinning direction.

We formulate our theory in terms of a pinning center aligned coordinate frame $(x',y',z')$. This new coordinate frame is related to the crystalline frame by a general three-dimensional rotation matrix $\hat{R}$. Hence, the system is equivalent to applying the corresponding rotation to the system in Fig. \ref{fig:setup}. The new coordinate system is then aligned such that the domain-wall normal is always in the $x'$ direction, the pinning direction is always in the $z'$ direction, and the vector between the two pinning centers is in the $y'$ direction. This allows for easier comparison between solutions. In the energy functional, this coordinate change is achieved by acting with the rotation matrix $\hat{R}$ on the anisotropy matrices $\hat{Q}^{\alpha\beta} \rightarrow \hat{R}^T \hat{Q}^{\alpha\beta} \hat{R}$.

\section{Magnetic Signatures and Numerical Solutions}
In an isotropic system, with translation invariance along the pinning direction $z^\prime$
and rotation invariance on the $\Omega$ plane,  it is sufficient to
consider only the $B_{z^\prime}$ component of the magnetic field. This can be obtained by simulating a 2D cross section of the superconductor with only the vector potential components $A_{x^\prime}$ and $A_{y^\prime}$.
However, the presence of anisotropies introduces an energetically preferred direction for the magnetic field, breaking the rotation symmetry. In addition, the magnetic field is, in general, no longer perpendicular to $\Omega$. It is, therefore, necessary to include the third vector potential component $A_{z^\prime}$. Note that this generalization is still compatible with translation invariance along the pinning direction.
The structure of the vector potential is $\vb{A} = \qty(A_{x^\prime}(x^\prime,y^\prime),A_{y^\prime}(x^\prime,y^\prime),A_{z^\prime}(x^\prime,y^\prime))$ and, consequently, the magnetic-field $\vb{B} = \qty(\partial_{y^\prime}A_{z^\prime},-\partial_{x^\prime} A_{z^\prime},\partial_{x^\prime}A_{y^\prime}-\partial_{y^\prime}A_{x^\prime})$.

\subsection{Rotation about the $z$ axis}
\begin{figure}
	\center
	\includegraphics[width=0.60\linewidth]{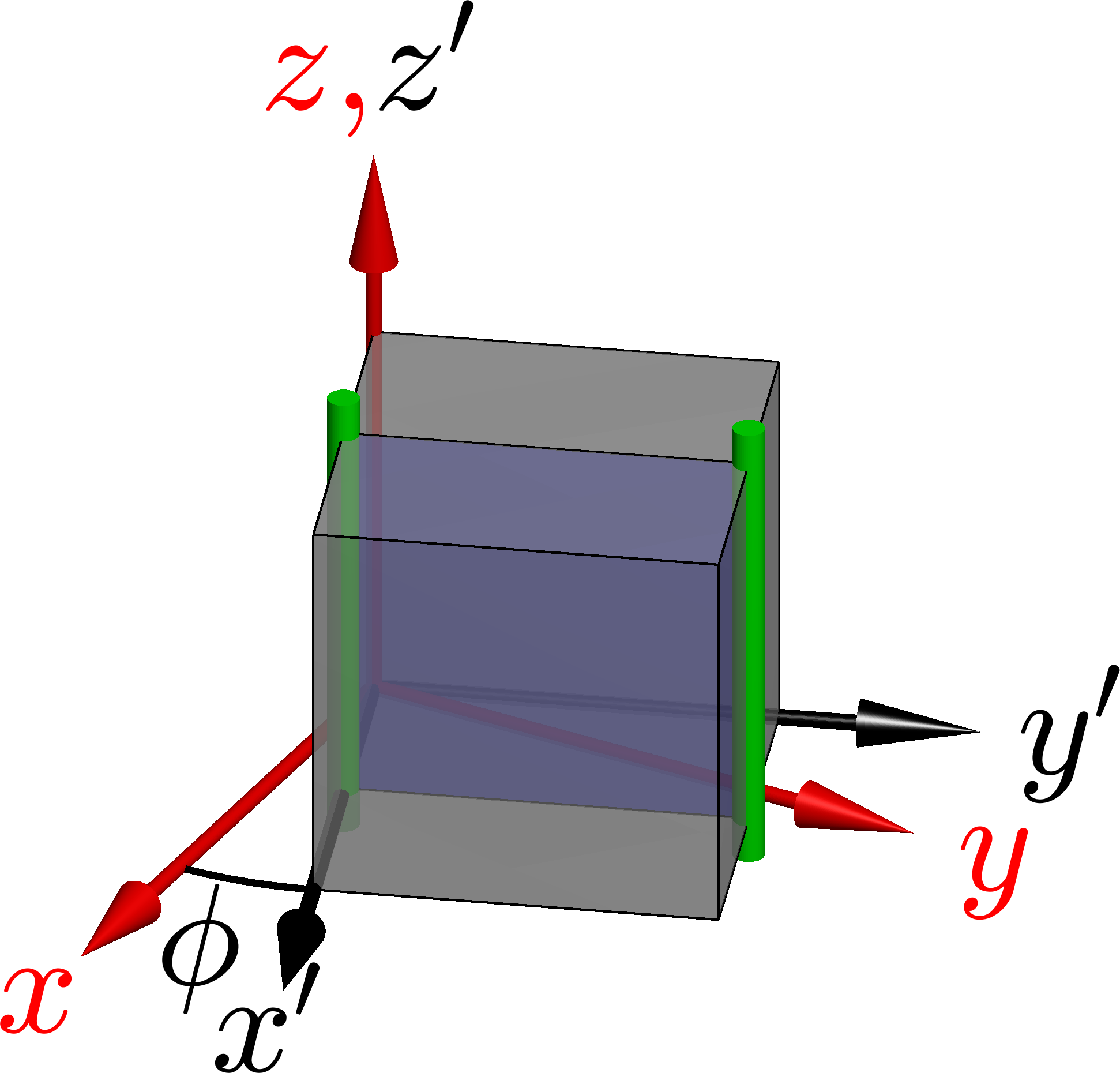}
	\caption{A rotation of the trivial setup in Fig. \ref{fig:setup} by $\phi$ about the $z$ axis, such that the domain-wall normal is not aligned with any of the crystalline axes $(x,y,z)$ which are drawn in red. The black axes represent the pinning-centers-aligned frame and are labeled by the primed coordinate set $(x^\prime,y^\prime,z^\prime)$ of which $x^\prime$ is always perpendicular to the domain wall. This coordinate frame configuration can be achieved by irradiation of the sample in a given direction or by cleaving the crystal, or creating dents on a surface of a small sample. }\label{fig:rotationExample}
\end{figure}

The first non-trivial orientation we consider, shown in Fig. \ref{fig:rotationExample}, is a rotation of the domain wall about the $z$ axis, corresponding to the rotation matrix,
\begin{equation}
\hat{R} =\left(
\begin{array}{ccc}
\cos \phi  & -\sin \phi  & 0 \\
\sin \phi  & \cos \phi  & 0 \\
0 & 0 & 1 \\
\end{array}
\right).
\label{eq:R_c}
\end{equation}
In the $s+is$ model, this rotation is a symmetry of the system due to the $SO(2)$ spatial rotation symmetry on the $xy$ plane.
In fact, independent of the value of the rotation angle $\phi$, all the couplings between the magnetic field and  the density gradients as well as the magnetic field and phase difference gradients cancel out (unless the domain wall interacts with a pinning center or inhomogeneity).
Therefore, for an $s+is$ superconductor, we do not have a \textit{bulk} magnetic signature for any $\phi$. 
For $s+id$ domain walls, the couplings no longer simplify, and we have a $\phi$-dependent \textit{bulk} magnetic field.

We have simulated the system described by the free energy functional in Eq. \eqref{eq:FreeEnergyDensity} in the domain $\Omega$ with angle of rotation $\phi = \pi/4$  using FreeFEM  \cite{FREEFEM} and a conjugate gradient flow energy-minimization method, with the results plotted in Fig. \ref{fig:2d_comparison}. The specific simulation parameters are reported in the Appendix.

The results demonstrate a marked difference between $s+is$ and $s+id$ domain walls. With the parameters we have selected, the matter field magnitudes give similar plots for both types, although, quantitatively, there are slight deviations due to couplings with the magnetic field. As predicted by the symmetries, $s+is$ domain walls exhibit no \textit{bulk} magnetic response. The localized magnetic field around the pinning centers is due to the non-convex geometry of the boundaries and is studied in detail in Ref. \cite{Garaud2014}. However, $s+id$ domain walls exhibit a strong spontaneous \textit{bulk} magnetic field, which extends along the entire length of the domain wall, instead of being localized at the pinning sites.
This field is characterized by a relatively strong magnitude, merely an order of magnitude smaller than the maximum magnetic field of a vortex in the same system and of the same order or stronger than the magnetic field resulting from impurity modulation.
This indicates that pinned domain walls can contribute strongly to spontaneous magnetic signatures in experiments \cite{grinenko2018emerging,Grinenko2017}. The origin of this magnetic signature can be identified in additional couplings among the magnetic field, gradients of phase difference, and matter field amplitudes, arising from the domain-wall normal vector not being aligned with any of the crystalline axes. This is an ideal orientation to consider experimentally as any $s+id$ domain wall will have a measurable magnetic response, compared with the $s+is$ case which has only a weaker localized response.
\begin{figure}
	\center
	\includegraphics[width=0.99\linewidth]{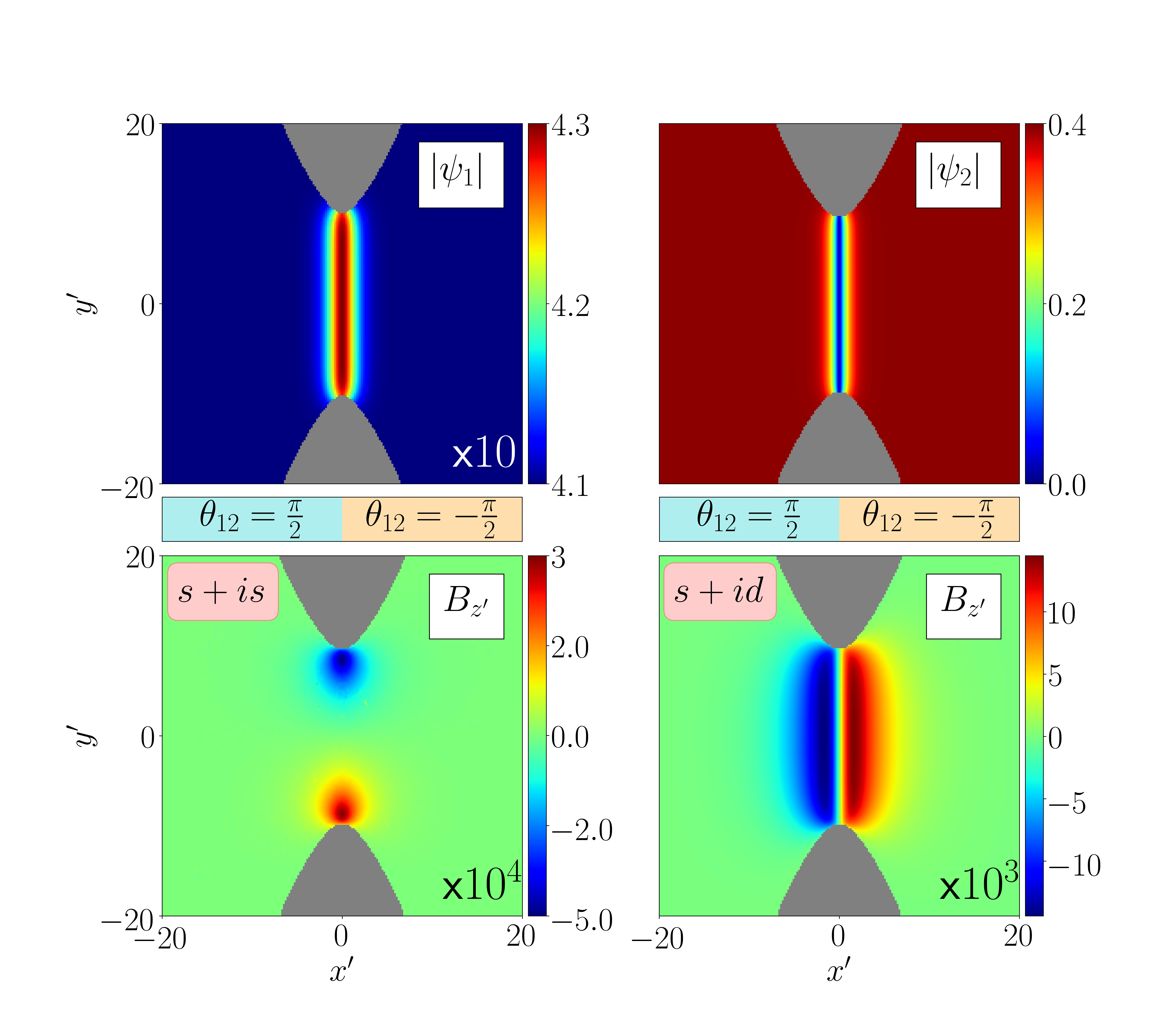}
		\caption{Order parameter's modulus $\abs{\psi_1}, \abs{\psi_2}$ and corresponding spontaneous magnetic field $B_{z^\prime}$ for $s+is$ and $s+id$ superconductors. We show a two-dimensional cross section of the sample, namely, the plane $\Omega$ in the pinning-centers-aligned coordinate frame. The pinning-centers-aligned axes are related to the crystalline axes through a rotation of $\phi=\pi/4$ around the $z$ crystalline axis, described by Eq. \eqref{eq:R_c}. The columnar pinning centers coincide with the gray areas. The two order parameters in both $s+is$ and $s+id$ superconductors are structurally similar, hence, we plot them only once. The phase difference is reported in the rectangular boxes, displaying a value $\theta_{12}= \pi / 2$ for $x^\prime<0$ and $\theta_{12}=-\pi / 2$ for $x^\prime>0$.
		In the magnetic-field plots, one can distinguish the qualitative difference among $s+is$ response, weak and localized around the pinning sites, and the $s+id$ response, stronger and extended for the entire length of the domain wall. Both magnetic fields directed along the $z^\prime$ direction. The calculation's parameters are reported in the Appendix.} \label{fig:2d_comparison}
\end{figure}

\subsection{Rotation about the $y$ axis}
More insight into the pairing symmetry can be obtained by considering a different orientation, namely, a rotation about the $y$ crystalline axis. This corresponds to the rotation matrix,
\begin{equation}
\hat{R}=
\left(
\begin{array}{ccc}
\cos \phi  & 0 & \sin \phi  \\
0 & 1 & 0 \\
-\sin \phi  & 0 & \cos \phi  \\
\end{array}
\right).
\label{R_b}
\end{equation}
\begin{figure}[!hbt]
	\center
	\hspace{0.25cm}
	\includegraphics[height=3.85cm]{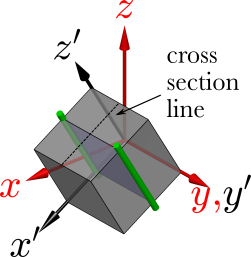}
	\hspace{0.25cm}
	\includegraphics[height=3.85cm]{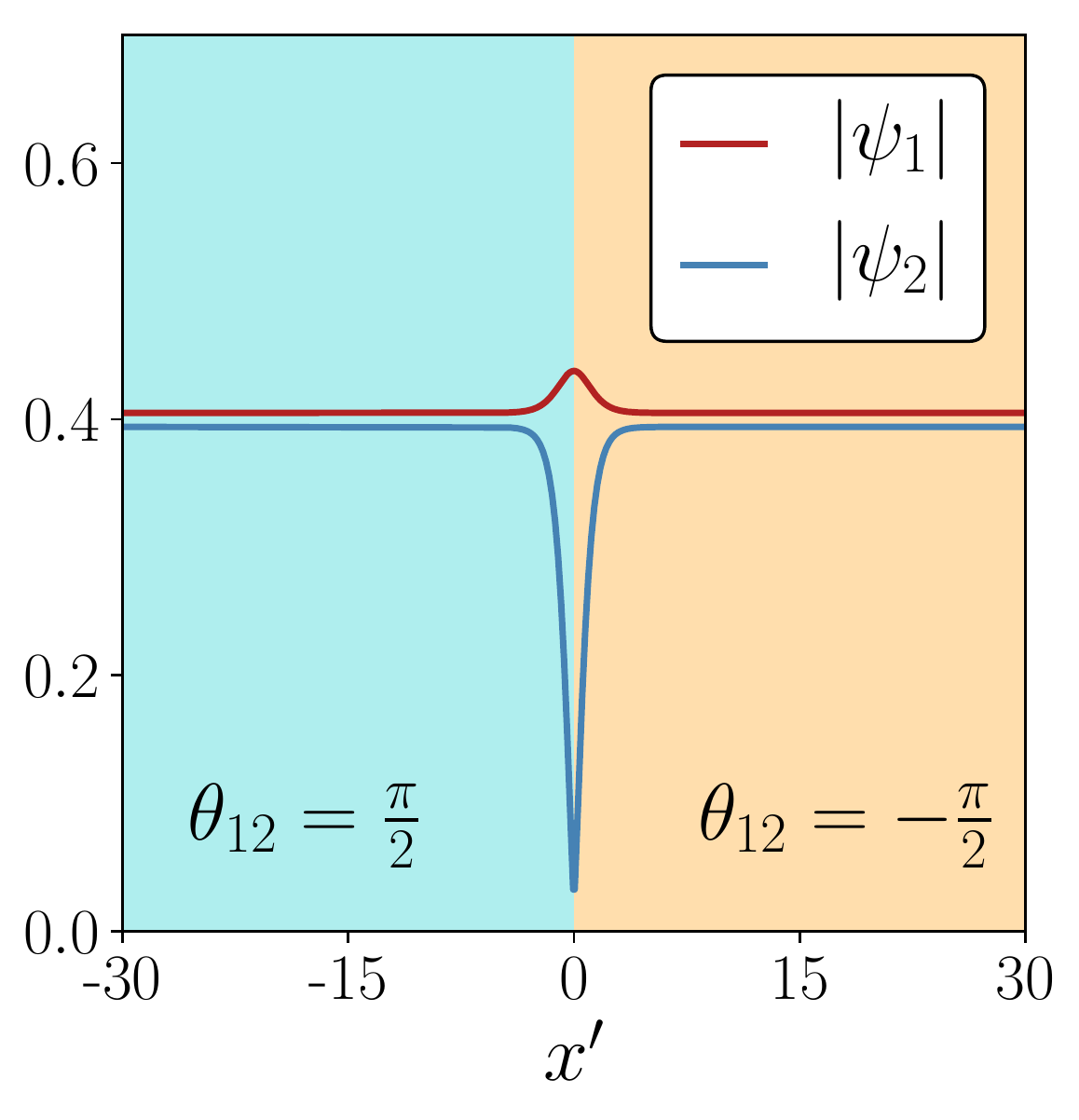} \\
	\vspace{0.5cm}
	\includegraphics[height=3.85cm]{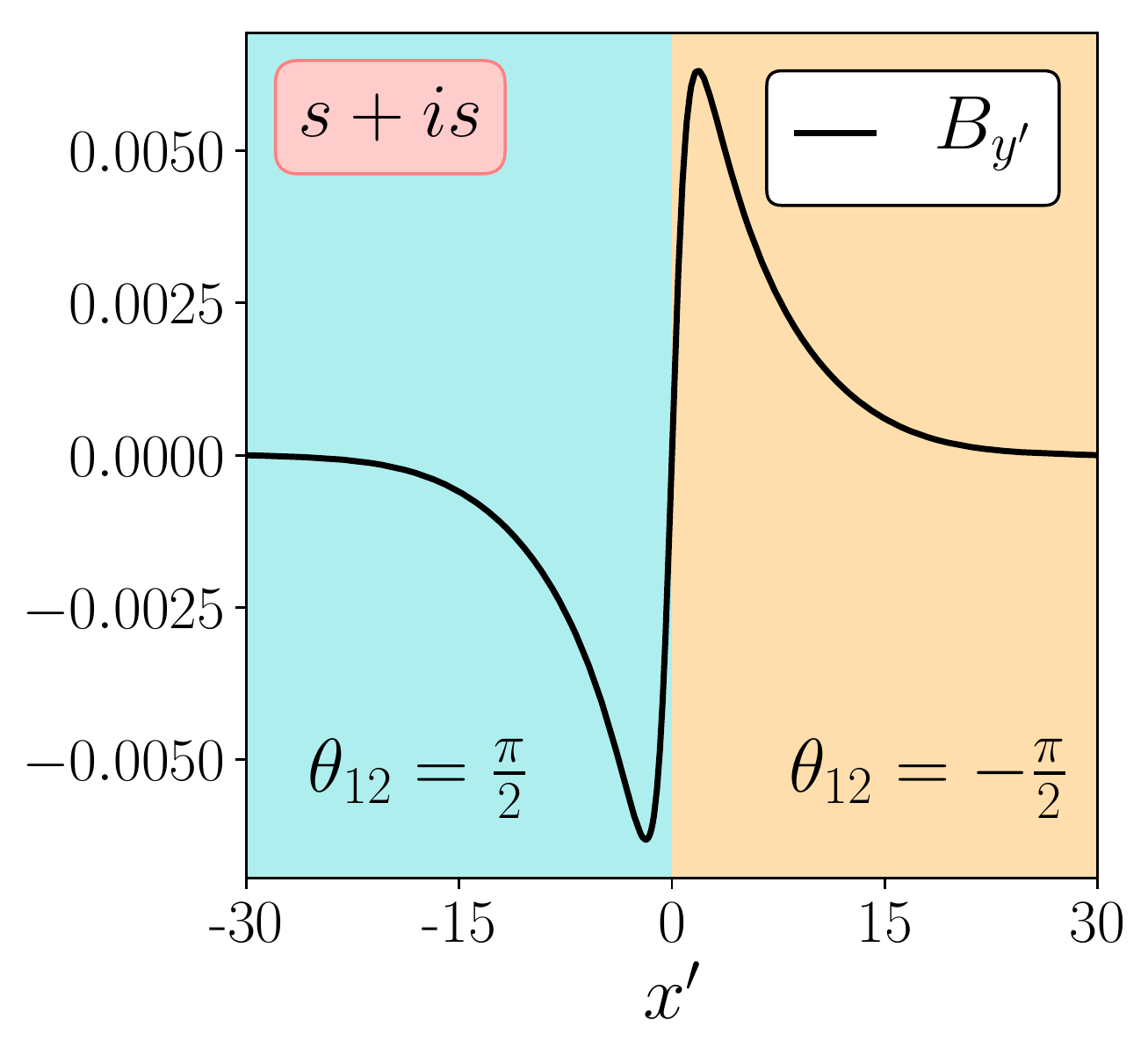}
	\includegraphics[height=3.85cm]{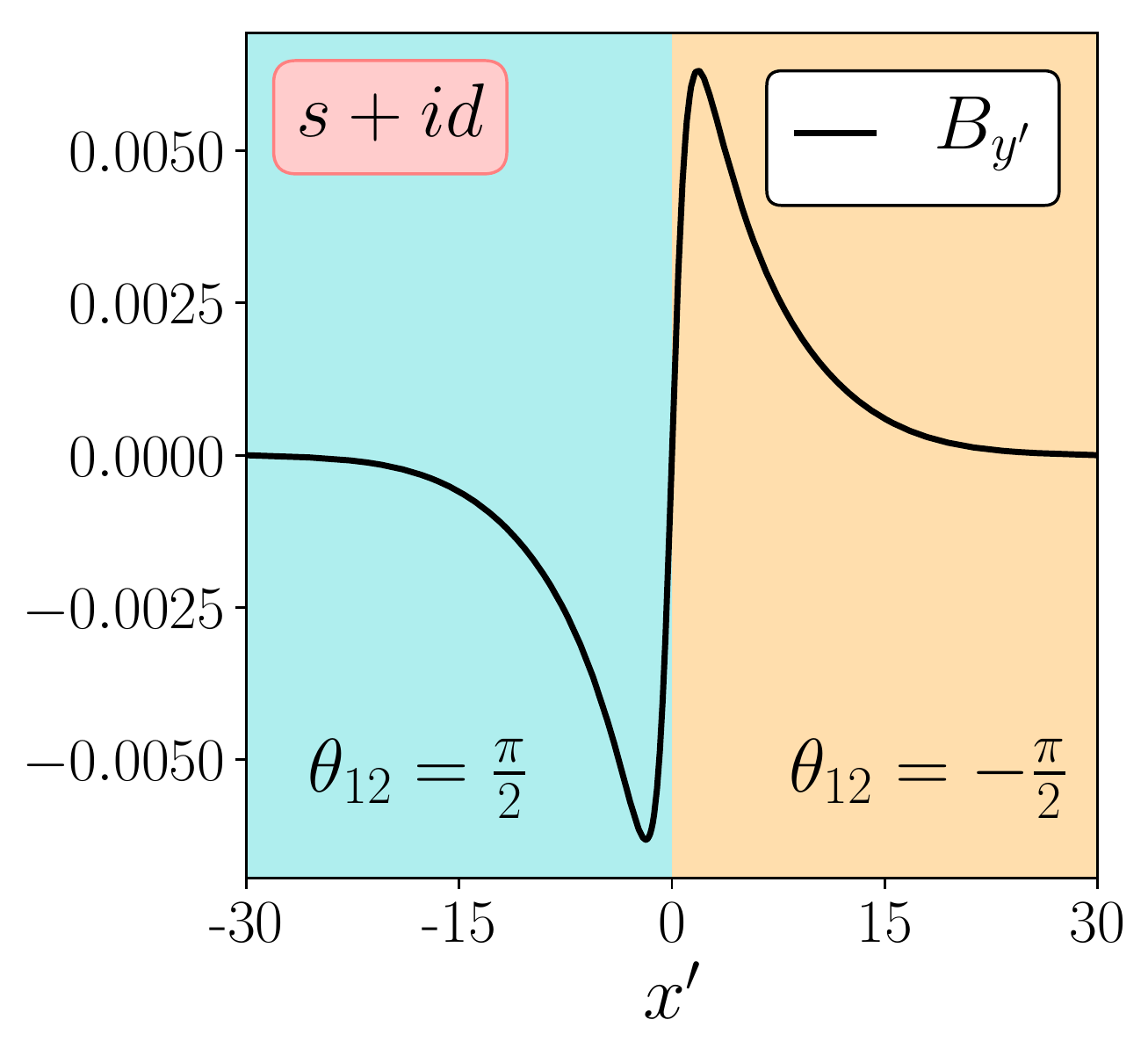}
	\caption{Simulations for the pinning setup rotated by $\phi = \pi/6$. About the $y$-crystalline axis. The top left panel displays the relative orientations of the crystalline coordinate frame and the pinning-centers-aligned coordinate frame. The dashed line indicates where we take the cross section of the fields, which is aligned with the normal to the domain wall and the $x^\prime$ axis. As in Fig. \ref{fig:2d_comparison}, the matter field modulus $\abs{\psi_{1}}$ and $\abs{\psi_{2}}$ behave in a similar way in both $s+is$ and $s+id$, therefore, we report them in a single panel on the top right. The phase difference is represented by the background color. Namely, the cyan (where $x' < 0$) indicates a phase difference value of $\theta_{12}={\pi} / {2}$ whereas orange (where $x' > 0$) indicates $\theta_{12}=-{\pi} / {2}$. 
	The two bottom panels show the spontaneous magnetic field for the $s+is$ sample on the left and the $s+id$ on the right. We can note that, under this system setup, both samples exhibit a quantitatively similar magnetic field, characterized by its extension throughout the entire length of the domain wall. It is substantially different if compared with Fig. \ref{fig:2d_comparison} for the $s+is$ case.}
		\label{fig:csLine}
\end{figure}
For this orientation, there is no simplification of the couplings for either the $s+is$ or the $s+id$ systems, and thus, we must simulate both numerically to study the structure of the spontaneous magnetic field. The results of this simulation for $\phi = \pi/6$ are plotted in Fig. \ref{fig:csLine}. The top left panel also displays the system setup we are considering. 
We note that, as we are purely interested in the \textit{bulk} response (the response far from the pinning sites), we have only plotted a cross section of the fields, taken along the $x^\prime$ axis, as highlighted in the top left panel of Fig. \ref{fig:csLine}. The resulting fields for $s+is$ and $s+id$ are very similar, demonstrating that both $s+is$ and $s+id$ domain walls exhibit spontaneous magnetic fields, which, in this case, is in the $y'$ direction. In fact, when the domain-wall normal is not aligned with any of the crystalline axes or on the $xy$ plane, $s+is$ domain walls will exhibit \textit{bulk} magnetic signatures which extend through the entire length of the domain wall.
As in Fig. \ref{fig:2d_comparison}, the order parameter magnitudes behave similarly in both states, hence, we report them only once. The phase difference value is plotted as background color. Cyan (where $x' < 0$) corresponds to a phase difference of $\theta_{12} = {\pi} / {2}$, whereas orange (where $x' > 0$) is associated with $\theta_{12} = -{\pi} / {2}$.

We note that the magnitude of the magnetic response strongly depends on the numerical value of the anisotropy matrices $\hat{Q}^{\alpha\beta}$. If we compare the magnitude of the maximal value of the domain wall's magnetic field with the maximal value of the magnetic field of a vortex in the same system, the magnetic field of the domain wall is ten times weaker compared to the vortex but of the same order of magnitude as the magnetic field generated by the impurity modulation considered in Ref. \cite{vadimov2018polarization}. 

\section{Complete Configuration Space}
In Fig. \ref{fig:csLine}, the magnetic-field strengths of $s+is$ and $s+id$ superconductors are very similar, and the magnetic-field direction is the same. In general, this is not the case, and the magnetic-field strength along with the magnetic-field direction will give information on the pairing symmetry as we have already seen in Fig. \ref{fig:2d_comparison}.

\begin{figure}
	\center
	\includegraphics[width=0.99\linewidth]{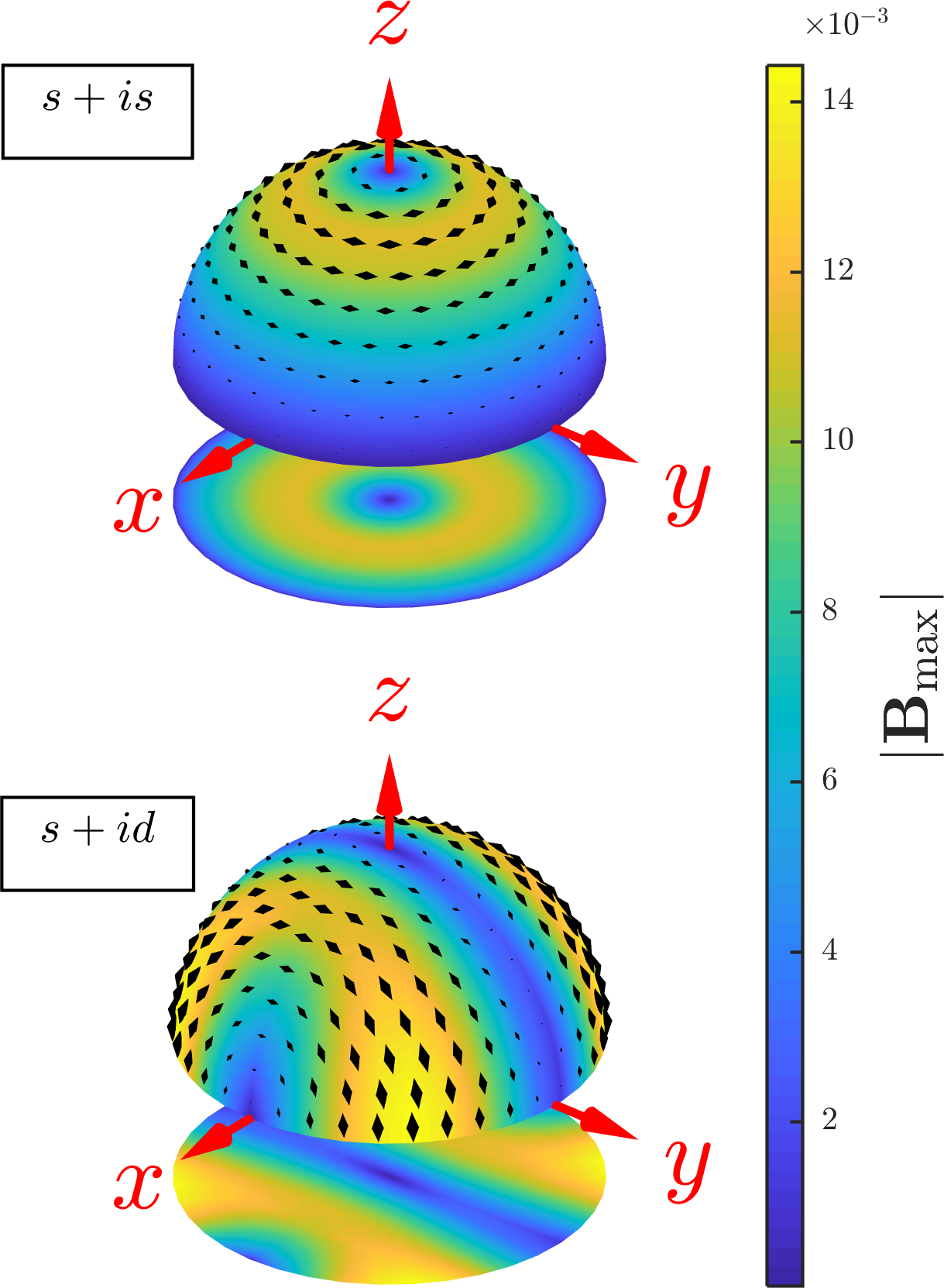}
	\caption{The maximum value and direction of the spontaneous magnetic field given a certain orientation of the domain-wall normal vector with respect to the crystalline coordinate frame, denoted by the coordinate set $(x,y,z)$. The upper image shows the magnetic field for the $s+is$ state and the lower image for the $s+id$ state. We can clearly note how different orientations of the domain-wall normal vector correspond to different spontaneous magnetic fields. The spontaneous magnetic field associated with the domain wall can be used to distinguish between the $s+is$ and the $s+id$ states since the magnetic field for the two states are only similar for restricted orientations of the domain wall. 
	}
	\label{fig:span}
\end{figure}

To consider all possible domain walls for our chosen parameters (reported in the Appendix), it is sufficient to consider all possible directions of the domain wall-normal $\vb{n}$. Due to the $C_2$ symmetry in the $z$ direction, we can consider just the directions of the normal in the upper hemisphere of a unit sphere. The results of considering all possible orientations of the domain wall are plotted in Fig. \ref{fig:span}. In this plot, each point indicates the domain wall-normal,  oriented from the origin to the point on a unit sphere in the crystalline coordinate system. The color of the point gives the maximum local strength of the magnetic field, whereas the arrow will give the unique magnetic field direction for both $s+is$ and $s+id$ systems. The arrow size scales with the values of $\abs{\vb B_\textrm{max}}$.
We see that the magnetic field's dependence on the orientation of the domain wall, relative to the crystalline axes, is markedly different for the 
$s+is$ and $s+id$ cases.
The easiest way to discriminate between the states can be seen from the color plot in Fig. \ref{fig:span}. It gives a clear demonstration of the symmetry on the basal ($xy$) plane, which, for $s+is$, is $SO(2)$ and, for $s+id$, is $C_2$. 
By computing the average of $|\vb{B}_{\rm{max}}|$ with respect to all possible domain-wall orientations, for both $s+is$ and $s+id$, we find that they are comparable in magnitude, since $\expval{|\vb{B}_{\rm{max}}|}_{s+is} /  \expval{|\vb{B}_{\rm{max}}|}_{s+id} \approx 2/3$. Hence, it is necessary to study the spontaneous magnetic field's dependence on the domain-wall orientation in order to determine the symmetry of the superconducting order parameter.
\section{Conclusion}

Superconducting states with spontaneously broken time reversal symmetry are of great current interest, however, identifying the type of BTRS order parameter is a notoriously difficult problem.
Recent experiments have reported the observation of broken time-reversal symmetry
in iron-based superconductors \cite{Grinenko2017,grinenko2018emerging}.
The evidence is based on spontaneous magnetic fields. \footnote{Thermodynamic evidence was also recently obtained Grinenko et. \cite{grinenko2018emerging}}
The leading candidates to explain these states are $s+is$ and $s+id$ pairings.

We have obtained solutions of domain walls in the $s+is$ and $s+id$ models of superconductors including the effects of anisotropies.
The solutions are obtained for different orientations of the domain walls relative to crystal axes,
and it is found that, in general, domain walls generate a spatially extended (\textit{bulk}) magnetic field in anisotropic superconductors.
For microscopically motivated \cite{garaud2017microscopically} parameters, the magnetic fields are substantial, only an order of magnitude smaller than that of a vortex and of the same order of magnitude as the magnetic signatures obtained from the impurity modulation studied in Ref. \cite{vadimov2018polarization}.
This demonstrates that the presence of domain walls should be an important contributing factor for the spontaneous magnetic field arising in BTRS states.
Superconducting samples naturally have defects, which means that domain walls will be spontaneously pinned. By irradiating the samples our goal is to increase the number of pinned domain walls with a specific orientation to enhance their contribution to the magnetic fields measured by means of muon spin rotation experiments. \cite{grinenko2018emerging,grinenko2018emerging}. 
Scanning SQUID and scanning Hall probes  \cite{moshchalkov2,moler1,moler2,gutierrez2012scanning} are also promising techniques to detect spontaneous magnetic fields. 
Importantly, the magnetic signatures in the $s+is$ and $s+id$ cases
are qualitatively and quantitatively different for different orientations of the domain wall. 
We presented a procedure where,
by a sequence of measurements with different orientations of fabricated  pinning centers, 
one can extract information about the symmetry of the order parameter from the
magnetic field generated by the domain wall.

\subsection*{Acknowledgments}
We thank M. Silaev and J. Garaud for many useful discussions.
The work of A.W., M.S. and T.W. is supported by the
U.K. Engineering and Physical Sciences Research Council
through a Ph.D. studentship (A.W.) and Grant No. EP/P024688/1. A.B., M.B. and E.B. were supported by
the Swedish Research Council Grants No. 642-2013-7837, No. 2018-03659,  No.
VR2016-06122, and the G{\"o}ran Gustafsson Foundation for Research
in Natural Sciences and Medicine.

\section{Appendix}
To obtain the (GL) expansion in Eq. \eqref{eq:FreeEnergyDensity}, we must consider the microscopic model for a clean superconductor with three overlapping bands at the Fermi level, which has been proposed to describe the BTRS superconducting state in iron-based superconductors \cite{Maiti2013, Marciani2013legett, StanevTesanovic}. The detailed derivation of the GL model with gradient terms
is given in Refs. \cite{garaud2017microscopically,Garaud2016}, and we adopt the results from these references.
Here, we briefly recall the key points in  that derivation. We start from the Ginzburg-Landau equations for a three-band superconductor. They are obtained by solving the Eilenberger equations and expanding the anomalous quasiclassical propagators in powers and gradients of the   gap functions $(\Delta_1,\Delta_2,\Delta_3)$. The Ginzburg-Landau equations for a three-band clean superconductor read
\begin{equation}\label{eq:GLmicro}
    \qty[\qty(G_0+\tau-\hat{\Lambda}^{-1})\vb{\Delta}]_\alpha=-\rho K_{ij}^{(\alpha)}\Pi_i\Pi_j\Delta_\alpha + \abs{\Delta_\alpha}^2\Delta_\alpha\,,
\end{equation}
where $\vb{\Delta}=(\Delta_1,\Delta_2,\Delta_3)$, $\tau = 1-T/T_c$, $\rho=\sum_n \pi T_c^3 \omega_n^{-3}\simeq 0.1$ and $G_0=\min(G_1,G_2)$. $G_1,G_2$ are the positive eigenvalues of $\hat{\Lambda}^{-1}$, obtained by inverting the coupling matrix in Eq. \eqref{eq:interactionMatrix}, i.e.,
\begin{equation}\label{eq:lamdaInverse}
     \hat{\Lambda}^{-1} =-\frac{1}{2\lambda^2\eta} 
    \mqty(\lambda^2 & -\lambda^2 & \lambda\eta \\ 
         -\lambda^2 &  \lambda^2 &-\lambda\eta \\ 
         -\lambda\eta & \lambda\eta & \eta^2)\, ,
\end{equation}
with 
\begin{equation}
    G_1 = \frac{1}{\eta},\qquad  G_2 = \frac{1}{4\lambda^2}\qty(\eta + \sqrt{\eta^2 + 8 \lambda^2})
\end{equation}
Since $\eta,\lambda>0$, one eigenvalue of $\hat{\Lambda}^{-1}$ is negative,  which implies that, in the Ginzburg-Landau expansion, we should retain two components. It is a rather common situation that, in a $N$-band superconductor, one cannot, in general, perform an expansion in $N$ small gaps and $N$ small gradient terms when the interband coupling is strong relative to intraband coupling \cite{silaev2012a}.
The tensor $K_{ij}^{(\alpha)}$ contains the information about the band anisotropy and is defined as
\begin{equation}\label{eq:Ktensors}
 K_{ij}^{(\alpha)} = \frac{\hbar^2\langle (\vb{v}^{(\alpha)}_F)_i(\vb{v}^{(\alpha)}_F)_j\rangle}{2T_c^2},
\end{equation}
where the average is taken over the $\alpha$th Fermi surface and $i,j$ indicate the spatial directions ($x,y,z$). 
Experimentally, it is challenging to determine these quantities, however, Refs. \cite{Exp-1,Exp-2} suggest that, for the majority of 122 iron pnictide materials, it makes sense to consider $K^{(\alpha)}_{xx}/K^{(\alpha)}_{zz} \in [1,5]$ as well as $K^{(\alpha)}_{yy}/K^{(\alpha)}_{zz} \in [1,5]$.

In  our paper, we consider a system with repulsive interband dominated pairing, where the matrix $\Lambda^{-1}$ in Eq. \eqref{eq:lamdaInverse} has only two positive eigenvalues. Hence, we follow the procedure in Refs. \cite{garaud2017microscopically,Garaud2016} to construct the effective two-component Ginzburg-Landau system of equations by applying the following transformation:
\begin{equation}\label{eq:transformation}
    \qty(\Delta_1,\Delta_2,\Delta_3) = \qty(\zeta\psi_2 - \psi_1, \zeta\psi_2 + \psi_1, \psi_2)\,,
\end{equation}
with $\zeta = \qty(\eta - \sqrt{\eta^2 + 8\lambda^2})/(4\lambda)$ and $\psi_1,\psi_2$ are the complex GL order parameters. By substituting Eq. \eqref{eq:transformation} into Eq. \eqref{eq:GLmicro}, we obtain the following system of equations (in the crystalline axes reference frame):
\begin{align}\label{eq:newGL1}
    a_1\psi_1 + b_{1}\abs{\psi_\alpha}^2\psi_1 + \gamma \abs{\psi_2}^2\psi_1  &+ \delta \psi_1^*\psi_2^2 = \\ \nonumber
    \rho \qty(K_{ii}^{(1)} + K_{ii}^{(2)})\Pi^2_i\psi_1 &+
    \rho \zeta\qty(K_{ii}^{(2)}-K_{ii}^{(1)})\Pi_i^2 \psi_2,\\
    \label{eq:newGL2}
    a_2\psi_2 + b_{2}\abs{\psi_2}^2\psi_2 + \gamma\abs{\psi_1}^2\psi_2 &+ \delta\psi_2^*\psi_1^2 = \\
    \nonumber
   \rho \qty[ \zeta^2(K_{ii}^{(1)} + K_{ii}^{(2)}) + K_{ii}^{(3)} ]\Pi^2_i\psi_2 &+ \rho \zeta\qty(K_{ii}^{(2)}-K_{ii}^{(1)})\Pi_i^2\psi_1\, ,
\end{align}
where we have introduced the parameters,
\begin{align}
	\nonumber
	&a_1     	  = -2\qty(G_0-G_1+\tau);\\
	&a_2          = -\qty(2\zeta^2+1)(G_0-G_2+\tau);\\
	&b_{1}    	  = 2; \qquad b_{2}= 1 + 2\zeta^4; \\\nonumber
	&\gamma       = 4\zeta^2; \quad \delta = 2\zeta^2\, . \\ \nonumber
\end{align}
The system of equations in Eqs. \eqref{eq:newGL1} and \eqref{eq:newGL2} can describe both $s+is$ and $s+id$ superconducting states, depending on the structures of the tensors in Eq. \eqref{eq:Ktensors}.  In the crystalline axes, the anisotropy tensors  $K_{ij}^{(\alpha)}$ are diagonal in their spatial components, and their symmetry requirements are reported in Table \ref{tab:conditions}.
\begin{table}[h]
 \begin{tabular}{c | c}
	\hline
	$s+is$ & $s+id$\\
	\hline
	\\
	$ K^{(1)}_{xx} = K^{(1)}_{yy} $ &$ K_{xx}^{(1)}=K_{yy}^{(2)}$\\
	$ K^{(2)}_{xx} = K^{(2)}_{yy} $ & $ K_{xx}^{(2)}=K_{yy}^{(1)}$\\
	$ K^{(3)}_{xx} = K^{(3)}_{yy} $ & $ K_{xx}^{(3)}=K_{yy}^{(3)}$\\
	\\
	\hline
\end{tabular}
\caption{Conditions on the elements of the anisotropy tensors to describe either the $s+is$ or the $s+id$ superconductor. }\label{tab:conditions}
\end{table}

Since in Eqs. \eqref{eq:newGL1} and \eqref{eq:newGL2} we only have combinations of $K_{ij}^{(\alpha)}$, we introduce the following tensors:
\begin{align}\label{eq:Qdefinitions}
\begin{split}
    Q^{11}_{ij} &= \rho \qty(K_{ij}^{(1)} + K_{ij}^{(2)})\\
    Q^{22}_{ij} &= \rho \qty[\zeta^2\qty(K_{ij}^{(1)} + K_{ij}^{(2)}) + K_{ij}^{(3)}]\\
    Q^{12}_{ij} &= \rho \zeta\qty(K_{ij}^{(1)} - K_{ij}^{(2)})
\end{split}
\end{align}
Hence, the difference between $s+is$ and $s+id$ states will be only in the structure of the tensor $Q^{12}_{ij}$.
The free-energy functional yielding the equations of motion Eqs. \eqref{eq:newGL1} and \eqref{eq:newGL2} is
\begin{align} \label{eq:FED}
	\begin{split}
	F &= \int \dd^3 x {\Bigg\{ \qty(\Pi_i\psi_\alpha)^* Q^{\alpha\beta}_{ij}\qty(\Pi_j\psi_\beta) + a_\alpha\abs{\psi_\alpha}^2 + \frac{b_\alpha}{2}\abs{\psi_\alpha}^4} \\ 
	&+\gamma\abs{\psi_1}^2 \abs{\psi_2}^2 +  \frac{\delta}{2}\qty(\psi_1^{*2}\psi_2^{2} + \psi_1^{2}\psi_2^{*2}) + \frac{\qty(\curl{\vb{A}})^2}{8\pi} \Bigg\}.
    \end{split}
\end{align}
which corresponds to Eq. \eqref{eq:FreeEnergyDensity} used in our paper.
In our simulations, we fixed $\eta = 5 $, $\lambda= 4.5 $ and $\tau=0.2$. To be in a BTRS regime, it is necessary to have $\eta/\lambda\sim 1$ and $\tau \in [0,0.3]$ as reported in Ref. \cite{garaud2017microscopically}. The choice of $K_{ij}^{(\alpha)}$ is reported in Table \ref{tab:Kcoefficients}.
\begin{table}[h]
 \begin{tabular}{c | c}
	\hline
	$s+is$ & $s+id$\\
	\hline
	\\
	$\hat{K}_1=\mqty(1.0 &0 &0 \\ 0 & 1.0 & 0 \\ 0 & 0 & 0.5)\quad $ & $\quad\hat{K}_1 =\mqty(1.0 &0 &0 \\ 0 & 1.5 & 0 \\ 0 & 0 & 0.5)$\\
	\\
	$\hat{K}_2=\mqty(1.5 &0 &0 \\ 0 & 1.5 & 0 \\ 0 & 0 & 0.3)\quad $  & $\quad \hat{K}_2 =\mqty(1.5 &0 &0 \\ 0 & 1.0 & 0 \\ 0 & 0 & 0.3)$\\
	\\
	$\hat{K}_3=\mqty(0.5 &0 &0 \\ 0 & 0.5 & 0 \\ 0 & 0 & 0.4)\quad $ & $\quad \hat{K}_3 =\mqty(0.5 &0 &0 \\ 0 & 0.5 & 0 \\ 0 & 0 & 0.4)$\\
	\\
	\hline
\end{tabular}
\caption{Anisotropy matrices $K_{ij}^{(\alpha)}$ in Eq. \eqref{eq:GLmicro} used in this paper, for both $s+is$ and $s+id$ systems. These matrices are written in the crystalline reference frame, in fact, they are diagonal in the spatial components. In this case, $\alpha\in\qty{1-3}$ since these matrices are directly related to the three microscopic bands.}\label{tab:Kcoefficients}.
\end{table} 

This choice of $K_{ij}^{(\alpha)}$ together with the definitions in Eq. \eqref{eq:Qdefinitions} yields to the $\hat{Q}^{\alpha\beta}$ matrices displayed in Table \ref{tab:Qcoefficients}.
\begin{table}[t]
 \begin{tabular}{c | c}
	\hline
	$s+is$ & $s+id$\\
	\hline
	\\
	$\hat{Q}^{11}\eqsim\mqty(0.25 &0 &0 \\ 0 & 0.25 & 0 \\ 0 & 0 & 0.08)$ & $\hat{Q}^{11}\eqsim\mqty(0.25 &0 &0 \\ 0 & 0.25 & 0 \\ 0 & 0 & 0.08)$\\
	\\
	$\hat{Q}^{22}\eqsim\mqty(0.11 &0 &0 \\ 0 & 0.11 & 0 \\ 0 & 0 & 0.06)$  & $\hat{Q}^{22}\eqsim\mqty(0.11 &0 &0 \\ 0 & 0.11 & 0 \\ 0 & 0 & 0.06)$\\
	\\
	$\hat{Q}^{12}\eqsim\mqty(0.024 &0 &0 \\ 0 & 0.024 & 0 \\ 0 & 0 & -0.01) $ & $\hat{Q}^{12}\eqsim\mqty(0.024 &0 &0 \\ 0 & -0.024 & 0 \\ 0 & 0 & -0.01)$\\
	\\
	\hline
\end{tabular}
\caption{Simulation parameters for anisotropy matrices $\hat{Q}^{\alpha\beta}$ for both $s+is$ and $s+id$ systems in the effective two-band Ginzburg-Landau model reported in Eq. \eqref{eq:FED}. These matrices are written in the crystalline reference frame, in fact, they are diagonal in the spatial components. In this case, $\alpha,\beta\in\qty{1,2}$. The values for $\hat{Q}^{\alpha\beta}$ are obtained from Eq. \eqref{eq:Qdefinitions}. }\label{tab:Qcoefficients}
\end{table} 
Finally, in the covariant derivative $\Pi_j=\partial_j+\textrm{i}qA_j$, we set $q=0.25$.
\bibliography{bibliography}

\begin{thebibliography}{37}%
\makeatletter
\providecommand \@ifxundefined [1]{%
 \@ifx{#1\undefined}
}%
\providecommand \@ifnum [1]{%
 \ifnum #1\expandafter \@firstoftwo
 \else \expandafter \@secondoftwo
 \fi
}%
\providecommand \@ifx [1]{%
 \ifx #1\expandafter \@firstoftwo
 \else \expandafter \@secondoftwo
 \fi
}%
\providecommand \natexlab [1]{#1}%
\providecommand \enquote  [1]{``#1''}%
\providecommand \bibnamefont  [1]{#1}%
\providecommand \bibfnamefont [1]{#1}%
\providecommand \citenamefont [1]{#1}%
\providecommand \href@noop [0]{\@secondoftwo}%
\providecommand \href [0]{\begingroup \@sanitize@url \@href}%
\providecommand \@href[1]{\@@startlink{#1}\@@href}%
\providecommand \@@href[1]{\endgroup#1\@@endlink}%
\providecommand \@sanitize@url [0]{\catcode `\\12\catcode `\$12\catcode
  `\&12\catcode `\#12\catcode `\^12\catcode `\_12\catcode `\%12\relax}%
\providecommand \@@startlink[1]{}%
\providecommand \@@endlink[0]{}%
\providecommand \url  [0]{\begingroup\@sanitize@url \@url }%
\providecommand \@url [1]{\endgroup\@href {#1}{\urlprefix }}%
\providecommand \urlprefix  [0]{URL }%
\providecommand \Eprint [0]{\href }%
\providecommand \doibase [0]{http://dx.doi.org/}%
\providecommand \selectlanguage [0]{\@gobble}%
\providecommand \bibinfo  [0]{\@secondoftwo}%
\providecommand \bibfield  [0]{\@secondoftwo}%
\providecommand \translation [1]{[#1]}%
\providecommand \BibitemOpen [0]{}%
\providecommand \bibitemStop [0]{}%
\providecommand \bibitemNoStop [0]{.\EOS\space}%
\providecommand \EOS [0]{\spacefactor3000\relax}%
\providecommand \BibitemShut  [1]{\csname bibitem#1\endcsname}%
\let\auto@bib@innerbib\@empty
\bibitem [{\citenamefont {Grinenko}\ \emph {et~al.}(2017)\citenamefont
  {Grinenko}, \citenamefont {Materne}, \citenamefont {Sarkar}, \citenamefont
  {Luetkens}, \citenamefont {Kihou}, \citenamefont {Lee}, \citenamefont
  {Akhmadaliev}, \citenamefont {Efremov}, \citenamefont {Drechsler},\ and\
  \citenamefont {Klauss}}]{Grinenko2017}%
  \BibitemOpen
  \bibfield  {author} {\bibinfo {author} {\bibfnamefont {V.}~\bibnamefont
  {Grinenko}}, \bibinfo {author} {\bibfnamefont {P.}~\bibnamefont {Materne}},
  \bibinfo {author} {\bibfnamefont {R.}~\bibnamefont {Sarkar}}, \bibinfo
  {author} {\bibfnamefont {H.}~\bibnamefont {Luetkens}}, \bibinfo {author}
  {\bibfnamefont {K.}~\bibnamefont {Kihou}}, \bibinfo {author} {\bibfnamefont
  {C.~H.}\ \bibnamefont {Lee}}, \bibinfo {author} {\bibfnamefont
  {S.}~\bibnamefont {Akhmadaliev}}, \bibinfo {author} {\bibfnamefont {D.~V.}\
  \bibnamefont {Efremov}}, \bibinfo {author} {\bibfnamefont {S.-L.}\
  \bibnamefont {Drechsler}}, \ and\ \bibinfo {author} {\bibfnamefont {H.-H.}\
  \bibnamefont {Klauss}},\ }\href
  {https://link.aps.org/doi/10.1103/PhysRevB.95.214511} {\bibfield  {journal}
  {\bibinfo  {journal} {Phys. Rev. B}\ }\textbf {\bibinfo {volume} {95}},\
  \bibinfo {pages} {214511} (\bibinfo {year} {2017})}\BibitemShut {NoStop}%
\bibitem [{\citenamefont {Grinenko}\ \emph {et~al.}(2018)\citenamefont
  {Grinenko}, \citenamefont {Sarkar}, \citenamefont {Kihou}, \citenamefont
  {Lee}, \citenamefont {Morozov}, \citenamefont {Aswartham}, \citenamefont
  {B{\"u}chner}, \citenamefont {Chekhonin}, \citenamefont {Skrotzki},
  \citenamefont {Nenkov} \emph {et~al.}}]{grinenko2018emerging}%
  \BibitemOpen
  \bibfield  {author} {\bibinfo {author} {\bibfnamefont {V.}~\bibnamefont
  {Grinenko}}, \bibinfo {author} {\bibfnamefont {R.}~\bibnamefont {Sarkar}},
  \bibinfo {author} {\bibfnamefont {K.}~\bibnamefont {Kihou}}, \bibinfo
  {author} {\bibfnamefont {C.}~\bibnamefont {Lee}}, \bibinfo {author}
  {\bibfnamefont {I.}~\bibnamefont {Morozov}}, \bibinfo {author} {\bibfnamefont
  {S.}~\bibnamefont {Aswartham}}, \bibinfo {author} {\bibfnamefont
  {B.}~\bibnamefont {B{\"u}chner}}, \bibinfo {author} {\bibfnamefont
  {P.}~\bibnamefont {Chekhonin}}, \bibinfo {author} {\bibfnamefont
  {W.}~\bibnamefont {Skrotzki}}, \bibinfo {author} {\bibfnamefont
  {K.}~\bibnamefont {Nenkov}},  \emph {et~al.},\ }\href@noop {} {\bibfield
  {journal} {\bibinfo  {journal} {arXiv preprint arXiv:1809.03610}\ } (\bibinfo
  {year} {2018})}\BibitemShut {NoStop}%
\bibitem [{\citenamefont {Stanev}\ and\ \citenamefont {Te\ifmmode
  \check{s}\else \v{s}\fi{}anovi\ifmmode~\acute{c}\else
  \'{c}\fi{}}(2010)}]{StanevTesanovic}%
  \BibitemOpen
  \bibfield  {author} {\bibinfo {author} {\bibfnamefont {V.}~\bibnamefont
  {Stanev}}\ and\ \bibinfo {author} {\bibfnamefont {Z.}~\bibnamefont
  {Te\ifmmode \check{s}\else \v{s}\fi{}anovi\ifmmode~\acute{c}\else
  \'{c}\fi{}}},\ }\href {\doibase 10.1103/PhysRevB.81.134522} {\bibfield
  {journal} {\bibinfo  {journal} {Phys. Rev. B}\ }\textbf {\bibinfo {volume}
  {81}},\ \bibinfo {pages} {134522} (\bibinfo {year} {2010})}\BibitemShut
  {NoStop}%
\bibitem [{\citenamefont {Lee}\ \emph {et~al.}(2009)\citenamefont {Lee},
  \citenamefont {Zhang},\ and\ \citenamefont {Wu}}]{Lee.Zhang.Wu:09}%
  \BibitemOpen
  \bibfield  {author} {\bibinfo {author} {\bibfnamefont {W.-C.}\ \bibnamefont
  {Lee}}, \bibinfo {author} {\bibfnamefont {S.-C.}\ \bibnamefont {Zhang}}, \
  and\ \bibinfo {author} {\bibfnamefont {C.}~\bibnamefont {Wu}},\ }\href
  {\doibase 10.1103/PhysRevLett.102.217002} {\bibfield  {journal} {\bibinfo
  {journal} {Phys. Rev. Lett.}\ }\textbf {\bibinfo {volume} {102}},\ \bibinfo
  {pages} {217002} (\bibinfo {year} {2009})}\BibitemShut {NoStop}%
\bibitem [{\citenamefont {Carlstr\"om}\ \emph {et~al.}(2011)\citenamefont
  {Carlstr\"om}, \citenamefont {Garaud},\ and\ \citenamefont
  {Babaev}}]{Carlstroem2011a}%
  \BibitemOpen
  \bibfield  {author} {\bibinfo {author} {\bibfnamefont {J.}~\bibnamefont
  {Carlstr\"om}}, \bibinfo {author} {\bibfnamefont {J.}~\bibnamefont {Garaud}},
  \ and\ \bibinfo {author} {\bibfnamefont {E.}~\bibnamefont {Babaev}},\ }\href
  {\doibase 10.1103/PhysRevB.84.134518} {\bibfield  {journal} {\bibinfo
  {journal} {Phys. Rev. B}\ }\textbf {\bibinfo {volume} {84}},\ \bibinfo
  {pages} {134518} (\bibinfo {year} {2011})}\BibitemShut {NoStop}%
\bibitem [{\citenamefont {Maiti}\ and\ \citenamefont
  {Chubukov}(2013)}]{Maiti2013}%
  \BibitemOpen
  \bibfield  {author} {\bibinfo {author} {\bibfnamefont {S.}~\bibnamefont
  {Maiti}}\ and\ \bibinfo {author} {\bibfnamefont {A.~V.}\ \bibnamefont
  {Chubukov}},\ }\href {https://link.aps.org/doi/10.1103/PhysRevB.87.144511}
  {\bibfield  {journal} {\bibinfo  {journal} {Phys. Rev. B}\ }\textbf {\bibinfo
  {volume} {87}},\ \bibinfo {pages} {144511} (\bibinfo {year}
  {2013})}\BibitemShut {NoStop}%
\bibitem [{\citenamefont {B\"oker}\ \emph {et~al.}(2017)\citenamefont
  {B\"oker}, \citenamefont {Volkov}, \citenamefont {Efetov},\ and\
  \citenamefont {Eremin}}]{Boeker2017}%
  \BibitemOpen
  \bibfield  {author} {\bibinfo {author} {\bibfnamefont {J.}~\bibnamefont
  {B\"oker}}, \bibinfo {author} {\bibfnamefont {P.~A.}\ \bibnamefont {Volkov}},
  \bibinfo {author} {\bibfnamefont {K.~B.}\ \bibnamefont {Efetov}}, \ and\
  \bibinfo {author} {\bibfnamefont {I.}~\bibnamefont {Eremin}},\ }\href
  {\doibase 10.1103/PhysRevB.96.014517} {\bibfield  {journal} {\bibinfo
  {journal} {Phys. Rev. B}\ }\textbf {\bibinfo {volume} {96}},\ \bibinfo
  {pages} {014517} (\bibinfo {year} {2017})}\BibitemShut {NoStop}%
\bibitem [{\citenamefont {Platt}\ \emph {et~al.}(2012)\citenamefont {Platt},
  \citenamefont {Thomale}, \citenamefont {Honerkamp}, \citenamefont {Zhang},\
  and\ \citenamefont {Hanke}}]{platt2012}%
  \BibitemOpen
  \bibfield  {author} {\bibinfo {author} {\bibfnamefont {C.}~\bibnamefont
  {Platt}}, \bibinfo {author} {\bibfnamefont {R.}~\bibnamefont {Thomale}},
  \bibinfo {author} {\bibfnamefont {C.}~\bibnamefont {Honerkamp}}, \bibinfo
  {author} {\bibfnamefont {S.-C.}\ \bibnamefont {Zhang}}, \ and\ \bibinfo
  {author} {\bibfnamefont {W.}~\bibnamefont {Hanke}},\ }\href {\doibase
  10.1103/PhysRevB.85.180502} {\bibfield  {journal} {\bibinfo  {journal} {Phys.
  Rev. B}\ }\textbf {\bibinfo {volume} {85}},\ \bibinfo {pages} {180502}
  (\bibinfo {year} {2012})}\BibitemShut {NoStop}%
\bibitem [{\citenamefont {Garaud}\ and\ \citenamefont
  {Babaev}(2014)}]{Garaud2014}%
  \BibitemOpen
  \bibfield  {author} {\bibinfo {author} {\bibfnamefont {J.}~\bibnamefont
  {Garaud}}\ and\ \bibinfo {author} {\bibfnamefont {E.}~\bibnamefont
  {Babaev}},\ }\href {\doibase 10.1103/PhysRevLett.112.017003} {\bibfield
  {journal} {\bibinfo  {journal} {Phys. Rev. Lett.}\ }\textbf {\bibinfo
  {volume} {112}},\ \bibinfo {pages} {017003} (\bibinfo {year}
  {2014})}\BibitemShut {NoStop}%
\bibitem [{\citenamefont {Maiti}\ \emph {et~al.}(2015)\citenamefont {Maiti},
  \citenamefont {Sigrist},\ and\ \citenamefont
  {Chubukov}}]{ChubukovMaitiSigrist}%
  \BibitemOpen
  \bibfield  {author} {\bibinfo {author} {\bibfnamefont {S.}~\bibnamefont
  {Maiti}}, \bibinfo {author} {\bibfnamefont {M.}~\bibnamefont {Sigrist}}, \
  and\ \bibinfo {author} {\bibfnamefont {A.}~\bibnamefont {Chubukov}},\ }\href
  {\doibase 10.1103/PhysRevB.91.161102} {\bibfield  {journal} {\bibinfo
  {journal} {Phys. Rev. B}\ }\textbf {\bibinfo {volume} {91}},\ \bibinfo
  {pages} {161102} (\bibinfo {year} {2015})}\BibitemShut {NoStop}%
\bibitem [{\citenamefont {Lin}\ \emph {et~al.}(2016)\citenamefont {Lin},
  \citenamefont {Maiti},\ and\ \citenamefont
  {Chubukov}}]{lin2016distinguishing}%
  \BibitemOpen
  \bibfield  {author} {\bibinfo {author} {\bibfnamefont {S.-Z.}\ \bibnamefont
  {Lin}}, \bibinfo {author} {\bibfnamefont {S.}~\bibnamefont {Maiti}}, \ and\
  \bibinfo {author} {\bibfnamefont {A.}~\bibnamefont {Chubukov}},\ }\href@noop
  {} {\bibfield  {journal} {\bibinfo  {journal} {Physical Review B}\ }\textbf
  {\bibinfo {volume} {94}},\ \bibinfo {pages} {064519} (\bibinfo {year}
  {2016})}\BibitemShut {NoStop}%
\bibitem [{\citenamefont {Silaev}\ \emph {et~al.}(2015)\citenamefont {Silaev},
  \citenamefont {Garaud},\ and\ \citenamefont {Babaev}}]{Silaev.Garaud.ea:15}%
  \BibitemOpen
  \bibfield  {author} {\bibinfo {author} {\bibfnamefont {M.}~\bibnamefont
  {Silaev}}, \bibinfo {author} {\bibfnamefont {J.}~\bibnamefont {Garaud}}, \
  and\ \bibinfo {author} {\bibfnamefont {E.}~\bibnamefont {Babaev}},\ }\href
  {\doibase 10.1103/PhysRevB.92.174510} {\bibfield  {journal} {\bibinfo
  {journal} {Phys. Rev. B}\ }\textbf {\bibinfo {volume} {92}},\ \bibinfo
  {pages} {174510} (\bibinfo {year} {2015})}\BibitemShut {NoStop}%
\bibitem [{\citenamefont {Garaud}\ \emph {et~al.}(2016)\citenamefont {Garaud},
  \citenamefont {Silaev},\ and\ \citenamefont {Babaev}}]{Garaud2016}%
  \BibitemOpen
  \bibfield  {author} {\bibinfo {author} {\bibfnamefont {J.}~\bibnamefont
  {Garaud}}, \bibinfo {author} {\bibfnamefont {M.}~\bibnamefont {Silaev}}, \
  and\ \bibinfo {author} {\bibfnamefont {E.}~\bibnamefont {Babaev}},\ }\href
  {https://link.aps.org/doi/10.1103/PhysRevLett.116.097002} {\bibfield
  {journal} {\bibinfo  {journal} {Phys. Rev. Lett.}\ }\textbf {\bibinfo
  {volume} {116}},\ \bibinfo {pages} {097002} (\bibinfo {year}
  {2016})}\BibitemShut {NoStop}%
\bibitem [{\citenamefont {Garaud}\ \emph {et~al.}(2018)\citenamefont {Garaud},
  \citenamefont {Corticelli}, \citenamefont {Silaev},\ and\ \citenamefont
  {Babaev}}]{garaud2018properties}%
  \BibitemOpen
  \bibfield  {author} {\bibinfo {author} {\bibfnamefont {J.}~\bibnamefont
  {Garaud}}, \bibinfo {author} {\bibfnamefont {A.}~\bibnamefont {Corticelli}},
  \bibinfo {author} {\bibfnamefont {M.}~\bibnamefont {Silaev}}, \ and\ \bibinfo
  {author} {\bibfnamefont {E.}~\bibnamefont {Babaev}},\ }\href@noop {}
  {\bibfield  {journal} {\bibinfo  {journal} {Physical Review B}\ }\textbf
  {\bibinfo {volume} {98}},\ \bibinfo {pages} {014520} (\bibinfo {year}
  {2018})}\BibitemShut {NoStop}%
\bibitem [{\citenamefont {Vadimov}\ and\ \citenamefont
  {Silaev}(2018)}]{vadimov2018polarization}%
  \BibitemOpen
  \bibfield  {author} {\bibinfo {author} {\bibfnamefont {V.~L.}\ \bibnamefont
  {Vadimov}}\ and\ \bibinfo {author} {\bibfnamefont {M.~A.}\ \bibnamefont
  {Silaev}},\ }\href {\doibase 10.1103/PhysRevB.98.104504} {\bibfield
  {journal} {\bibinfo  {journal} {Phys. Rev. B}\ }\textbf {\bibinfo {volume}
  {98}},\ \bibinfo {pages} {104504} (\bibinfo {year} {2018})}\BibitemShut
  {NoStop}%
\bibitem [{\citenamefont {Silaev}\ and\ \citenamefont
  {Babaev}(2013)}]{Silaev.Babaev:13}%
  \BibitemOpen
  \bibfield  {author} {\bibinfo {author} {\bibfnamefont {M.}~\bibnamefont
  {Silaev}}\ and\ \bibinfo {author} {\bibfnamefont {E.}~\bibnamefont
  {Babaev}},\ }\href {\doibase 10.1103/PhysRevB.88.220504} {\bibfield
  {journal} {\bibinfo  {journal} {Phys. Rev. B}\ }\textbf {\bibinfo {volume}
  {88}},\ \bibinfo {pages} {220504} (\bibinfo {year} {2013})}\BibitemShut
  {NoStop}%
\bibitem [{\citenamefont {Lin}\ and\ \citenamefont {Hu}(2012)}]{Lin2012}%
  \BibitemOpen
  \bibfield  {author} {\bibinfo {author} {\bibfnamefont {S.-Z.}\ \bibnamefont
  {Lin}}\ and\ \bibinfo {author} {\bibfnamefont {X.}~\bibnamefont {Hu}},\
  }\href {\doibase 10.1103/PhysRevLett.108.177005} {\bibfield  {journal}
  {\bibinfo  {journal} {Phys. Rev. Lett.}\ }\textbf {\bibinfo {volume} {108}},\
  \bibinfo {pages} {177005} (\bibinfo {year} {2012})}\BibitemShut {NoStop}%
\bibitem [{\citenamefont {Marciani}\ \emph {et~al.}(2013)\citenamefont
  {Marciani}, \citenamefont {Fanfarillo}, \citenamefont {Castellani},\ and\
  \citenamefont {Benfatto}}]{marciani2013legett}%
  \BibitemOpen
  \bibfield  {author} {\bibinfo {author} {\bibfnamefont {M.}~\bibnamefont
  {Marciani}}, \bibinfo {author} {\bibfnamefont {L.}~\bibnamefont
  {Fanfarillo}}, \bibinfo {author} {\bibfnamefont {C.}~\bibnamefont
  {Castellani}}, \ and\ \bibinfo {author} {\bibfnamefont {L.}~\bibnamefont
  {Benfatto}},\ }\href {\doibase 10.1103/PhysRevB.88.214508} {\bibfield
  {journal} {\bibinfo  {journal} {Phys. Rev. B}\ }\textbf {\bibinfo {volume}
  {88}},\ \bibinfo {pages} {214508} (\bibinfo {year} {2013})}\BibitemShut
  {NoStop}%
\bibitem [{\citenamefont {Garaud}\ \emph {et~al.}(2011)\citenamefont {Garaud},
  \citenamefont {Carlstr\"om},\ and\ \citenamefont {Babaev}}]{Garaud2011}%
  \BibitemOpen
  \bibfield  {author} {\bibinfo {author} {\bibfnamefont {J.}~\bibnamefont
  {Garaud}}, \bibinfo {author} {\bibfnamefont {J.}~\bibnamefont {Carlstr\"om}},
  \ and\ \bibinfo {author} {\bibfnamefont {E.}~\bibnamefont {Babaev}},\ }\href
  {\doibase 10.1103/PhysRevLett.107.197001} {\bibfield  {journal} {\bibinfo
  {journal} {Phys. Rev. Lett.}\ }\textbf {\bibinfo {volume} {107}},\ \bibinfo
  {pages} {197001} (\bibinfo {year} {2011})}\BibitemShut {NoStop}%
\bibitem [{\citenamefont {Garaud}\ \emph {et~al.}(2013)\citenamefont {Garaud},
  \citenamefont {Carlstr\"om}, \citenamefont {Babaev},\ and\ \citenamefont
  {Speight}}]{Garaud2013}%
  \BibitemOpen
  \bibfield  {author} {\bibinfo {author} {\bibfnamefont {J.}~\bibnamefont
  {Garaud}}, \bibinfo {author} {\bibfnamefont {J.}~\bibnamefont {Carlstr\"om}},
  \bibinfo {author} {\bibfnamefont {E.}~\bibnamefont {Babaev}}, \ and\ \bibinfo
  {author} {\bibfnamefont {M.}~\bibnamefont {Speight}},\ }\href {\doibase
  10.1103/PhysRevB.87.014507} {\bibfield  {journal} {\bibinfo  {journal} {Phys.
  Rev. B}\ }\textbf {\bibinfo {volume} {87}},\ \bibinfo {pages} {014507}
  (\bibinfo {year} {2013})}\BibitemShut {NoStop}%
\bibitem [{\citenamefont {Hirschfeld}\ \emph {et~al.}(2015)\citenamefont
  {Hirschfeld}, \citenamefont {Altenfeld}, \citenamefont {Eremin},\ and\
  \citenamefont {Mazin}}]{Hirschfeld2015}%
  \BibitemOpen
  \bibfield  {author} {\bibinfo {author} {\bibfnamefont {P.~J.}\ \bibnamefont
  {Hirschfeld}}, \bibinfo {author} {\bibfnamefont {D.}~\bibnamefont
  {Altenfeld}}, \bibinfo {author} {\bibfnamefont {I.}~\bibnamefont {Eremin}}, \
  and\ \bibinfo {author} {\bibfnamefont {I.~I.}\ \bibnamefont {Mazin}},\ }\href
  {\doibase 10.1103/PhysRevB.92.184513} {\bibfield  {journal} {\bibinfo
  {journal} {Phys. Rev. B}\ }\textbf {\bibinfo {volume} {92}},\ \bibinfo
  {pages} {184513} (\bibinfo {year} {2015})}\BibitemShut {NoStop}%
\bibitem [{\citenamefont {Bj\"ornsson}\ \emph {et~al.}(2005)\citenamefont
  {Bj\"ornsson}, \citenamefont {Maeno}, \citenamefont {Huber},\ and\
  \citenamefont {Moler}}]{moler1}%
  \BibitemOpen
  \bibfield  {author} {\bibinfo {author} {\bibfnamefont {P.~G.}\ \bibnamefont
  {Bj\"ornsson}}, \bibinfo {author} {\bibfnamefont {Y.}~\bibnamefont {Maeno}},
  \bibinfo {author} {\bibfnamefont {M.~E.}\ \bibnamefont {Huber}}, \ and\
  \bibinfo {author} {\bibfnamefont {K.~A.}\ \bibnamefont {Moler}},\ }\href
  {\doibase 10.1103/PhysRevB.72.012504} {\bibfield  {journal} {\bibinfo
  {journal} {Phys. Rev. B}\ }\textbf {\bibinfo {volume} {72}},\ \bibinfo
  {pages} {012504} (\bibinfo {year} {2005})}\BibitemShut {NoStop}%
\bibitem [{\citenamefont {Hicks}\ \emph {et~al.}(2010)\citenamefont {Hicks},
  \citenamefont {Kirtley}, \citenamefont {Lippman}, \citenamefont {Koshnick},
  \citenamefont {Huber}, \citenamefont {Maeno}, \citenamefont {Yuhasz},
  \citenamefont {Maple},\ and\ \citenamefont {Moler}}]{moler2}%
  \BibitemOpen
  \bibfield  {author} {\bibinfo {author} {\bibfnamefont {C.~W.}\ \bibnamefont
  {Hicks}}, \bibinfo {author} {\bibfnamefont {J.~R.}\ \bibnamefont {Kirtley}},
  \bibinfo {author} {\bibfnamefont {T.~M.}\ \bibnamefont {Lippman}}, \bibinfo
  {author} {\bibfnamefont {N.~C.}\ \bibnamefont {Koshnick}}, \bibinfo {author}
  {\bibfnamefont {M.~E.}\ \bibnamefont {Huber}}, \bibinfo {author}
  {\bibfnamefont {Y.}~\bibnamefont {Maeno}}, \bibinfo {author} {\bibfnamefont
  {W.~M.}\ \bibnamefont {Yuhasz}}, \bibinfo {author} {\bibfnamefont {M.~B.}\
  \bibnamefont {Maple}}, \ and\ \bibinfo {author} {\bibfnamefont {K.~A.}\
  \bibnamefont {Moler}},\ }\href {\doibase 10.1103/PhysRevB.81.214501}
  {\bibfield  {journal} {\bibinfo  {journal} {Phys. Rev. B}\ }\textbf {\bibinfo
  {volume} {81}},\ \bibinfo {pages} {214501} (\bibinfo {year}
  {2010})}\BibitemShut {NoStop}%
\bibitem [{\citenamefont {{Nishio}}\ \emph {et~al.}(2010)\citenamefont
  {{Nishio}}, \citenamefont {{Dao}}, \citenamefont {{Chen}}, \citenamefont
  {{Chibotaru}}, \citenamefont {{Kadowaki}},\ and\ \citenamefont
  {{Moshchalkov}}}]{moshchalkov2}%
  \BibitemOpen
  \bibfield  {author} {\bibinfo {author} {\bibfnamefont {T.}~\bibnamefont
  {{Nishio}}}, \bibinfo {author} {\bibfnamefont {V.~H.}\ \bibnamefont {{Dao}}},
  \bibinfo {author} {\bibfnamefont {Q.}~\bibnamefont {{Chen}}}, \bibinfo
  {author} {\bibfnamefont {L.~F.}\ \bibnamefont {{Chibotaru}}}, \bibinfo
  {author} {\bibfnamefont {K.}~\bibnamefont {{Kadowaki}}}, \ and\ \bibinfo
  {author} {\bibfnamefont {V.~V.}\ \bibnamefont {{Moshchalkov}}},\ }\href
  {\doibase 10.1103/PhysRevB.81.020506} {\bibfield  {journal} {\bibinfo
  {journal} {\prb}\ }\textbf {\bibinfo {volume} {81}},\ \bibinfo {eid} {020506}
  (\bibinfo {year} {2010})}\BibitemShut {NoStop}%
\bibitem [{\citenamefont {Gutierrez}\ \emph {et~al.}(2012)\citenamefont
  {Gutierrez}, \citenamefont {Raes}, \citenamefont {Silhanek}, \citenamefont
  {Li}, \citenamefont {Zhigadlo}, \citenamefont {Karpinski}, \citenamefont
  {Tempere},\ and\ \citenamefont {Moshchalkov}}]{gutierrez2012scanning}%
  \BibitemOpen
  \bibfield  {author} {\bibinfo {author} {\bibfnamefont {J.}~\bibnamefont
  {Gutierrez}}, \bibinfo {author} {\bibfnamefont {B.}~\bibnamefont {Raes}},
  \bibinfo {author} {\bibfnamefont {A.}~\bibnamefont {Silhanek}}, \bibinfo
  {author} {\bibfnamefont {L.}~\bibnamefont {Li}}, \bibinfo {author}
  {\bibfnamefont {N.}~\bibnamefont {Zhigadlo}}, \bibinfo {author}
  {\bibfnamefont {J.}~\bibnamefont {Karpinski}}, \bibinfo {author}
  {\bibfnamefont {J.}~\bibnamefont {Tempere}}, \ and\ \bibinfo {author}
  {\bibfnamefont {V.}~\bibnamefont {Moshchalkov}},\ }\href@noop {} {\bibfield
  {journal} {\bibinfo  {journal} {Physical Review B}\ }\textbf {\bibinfo
  {volume} {85}},\ \bibinfo {pages} {094511} (\bibinfo {year}
  {2012})}\BibitemShut {NoStop}%
\bibitem [{\citenamefont {Silaev}\ \emph {et~al.}(2018)\citenamefont {Silaev},
  \citenamefont {Winyard},\ and\ \citenamefont {Babaev}}]{silaev2017non}%
  \BibitemOpen
  \bibfield  {author} {\bibinfo {author} {\bibfnamefont {M.}~\bibnamefont
  {Silaev}}, \bibinfo {author} {\bibfnamefont {T.}~\bibnamefont {Winyard}}, \
  and\ \bibinfo {author} {\bibfnamefont {E.}~\bibnamefont {Babaev}},\ }\href
  {\doibase 10.1103/PhysRevB.97.174504} {\bibfield  {journal} {\bibinfo
  {journal} {Phys. Rev. B}\ }\textbf {\bibinfo {volume} {97}},\ \bibinfo
  {pages} {174504} (\bibinfo {year} {2018})}\BibitemShut {NoStop}%
\bibitem [{\citenamefont {Winyard}\ \emph
  {et~al.}(2019{\natexlab{a}})\citenamefont {Winyard}, \citenamefont {Silaev},\
  and\ \citenamefont {Babaev}}]{winyard2018}%
  \BibitemOpen
  \bibfield  {author} {\bibinfo {author} {\bibfnamefont {T.}~\bibnamefont
  {Winyard}}, \bibinfo {author} {\bibfnamefont {M.}~\bibnamefont {Silaev}}, \
  and\ \bibinfo {author} {\bibfnamefont {E.}~\bibnamefont {Babaev}},\ }\href
  {\doibase 10.1103/PhysRevB.99.064509} {\bibfield  {journal} {\bibinfo
  {journal} {Phys. Rev. B}\ }\textbf {\bibinfo {volume} {99}},\ \bibinfo
  {pages} {064509} (\bibinfo {year} {2019}{\natexlab{a}})}\BibitemShut
  {NoStop}%
\bibitem [{\citenamefont {Winyard}\ \emph
  {et~al.}(2019{\natexlab{b}})\citenamefont {Winyard}, \citenamefont {Silaev},\
  and\ \citenamefont {Babaev}}]{winyard2018skyrmion}%
  \BibitemOpen
  \bibfield  {author} {\bibinfo {author} {\bibfnamefont {T.}~\bibnamefont
  {Winyard}}, \bibinfo {author} {\bibfnamefont {M.}~\bibnamefont {Silaev}}, \
  and\ \bibinfo {author} {\bibfnamefont {E.}~\bibnamefont {Babaev}},\ }\href
  {\doibase 10.1103/PhysRevB.99.024501} {\bibfield  {journal} {\bibinfo
  {journal} {Phys. Rev. B}\ }\textbf {\bibinfo {volume} {99}},\ \bibinfo
  {pages} {024501} (\bibinfo {year} {2019}{\natexlab{b}})}\BibitemShut
  {NoStop}%
\bibitem [{\citenamefont {Garaud}\ \emph {et~al.}(2017)\citenamefont {Garaud},
  \citenamefont {Silaev},\ and\ \citenamefont
  {Babaev}}]{garaud2017microscopically}%
  \BibitemOpen
  \bibfield  {author} {\bibinfo {author} {\bibfnamefont {J.}~\bibnamefont
  {Garaud}}, \bibinfo {author} {\bibfnamefont {M.}~\bibnamefont {Silaev}}, \
  and\ \bibinfo {author} {\bibfnamefont {E.}~\bibnamefont {Babaev}},\
  }\href@noop {} {\bibfield  {journal} {\bibinfo  {journal} {Physica C:
  Superconductivity and its Applications}\ }\textbf {\bibinfo {volume} {533}},\
  \bibinfo {pages} {63} (\bibinfo {year} {2017})}\BibitemShut {NoStop}%
\bibitem [{\citenamefont {Speight}\ \emph {et~al.}(2019)\citenamefont
  {Speight}, \citenamefont {Winyard},\ and\ \citenamefont
  {Babaev}}]{speight2019chiral}%
  \BibitemOpen
  \bibfield  {author} {\bibinfo {author} {\bibfnamefont {M.}~\bibnamefont
  {Speight}}, \bibinfo {author} {\bibfnamefont {T.}~\bibnamefont {Winyard}}, \
  and\ \bibinfo {author} {\bibfnamefont {E.}~\bibnamefont {Babaev}},\
  }\href@noop {} {\bibfield  {journal} {\bibinfo  {journal} {arXiv preprint
  arXiv:1905.07296}\ } (\bibinfo {year} {2019})}\BibitemShut {NoStop}%
\bibitem [{\citenamefont {Espinosa-Arronte}\ \emph {et~al.}(2007)\citenamefont
  {Espinosa-Arronte}, \citenamefont {Andersson}, \citenamefont {van~der Beek},
  \citenamefont {Nikolaou}, \citenamefont {Lidmar},\ and\ \citenamefont
  {Wallin}}]{PhysRevB.75.100504}%
  \BibitemOpen
  \bibfield  {author} {\bibinfo {author} {\bibfnamefont {B.}~\bibnamefont
  {Espinosa-Arronte}}, \bibinfo {author} {\bibfnamefont {M.}~\bibnamefont
  {Andersson}}, \bibinfo {author} {\bibfnamefont {C.~J.}\ \bibnamefont {van~der
  Beek}}, \bibinfo {author} {\bibfnamefont {M.}~\bibnamefont {Nikolaou}},
  \bibinfo {author} {\bibfnamefont {J.}~\bibnamefont {Lidmar}}, \ and\ \bibinfo
  {author} {\bibfnamefont {M.}~\bibnamefont {Wallin}},\ }\href {\doibase
  10.1103/PhysRevB.75.100504} {\bibfield  {journal} {\bibinfo  {journal} {Phys.
  Rev. B}\ }\textbf {\bibinfo {volume} {75}},\ \bibinfo {pages} {100504}
  (\bibinfo {year} {2007})}\BibitemShut {NoStop}%
\bibitem [{\citenamefont {Kibble}(1976)}]{kibble}%
  \BibitemOpen
  \bibfield  {author} {\bibinfo {author} {\bibfnamefont {T.~W.~B.}\
  \bibnamefont {Kibble}},\ }\href {\doibase 10.1088/0305-4470/9/8/029}
  {\bibfield  {journal} {\bibinfo  {journal} {Journal of Physics A:
  Mathematical and General}\ }\textbf {\bibinfo {volume} {9}},\ \bibinfo
  {pages} {1387} (\bibinfo {year} {1976})}\BibitemShut {NoStop}%
\bibitem [{\citenamefont {Zurek}(1985)}]{zurek}%
  \BibitemOpen
  \bibfield  {author} {\bibinfo {author} {\bibfnamefont {W.~H.}\ \bibnamefont
  {Zurek}},\ }\href {\doibase 10.1038/317505a0} {\bibfield  {journal} {\bibinfo
   {journal} {Nature}\ }\textbf {\bibinfo {volume} {317}},\ \bibinfo {pages}
  {505} (\bibinfo {year} {1985})}\BibitemShut {NoStop}%
\bibitem [{\citenamefont {Hecht}(2012)}]{FREEFEM}%
  \BibitemOpen
  \bibfield  {author} {\bibinfo {author} {\bibfnamefont {F.}~\bibnamefont
  {Hecht}},\ }\href {https://freefem.org/} {\bibfield  {journal} {\bibinfo
  {journal} {J. Numer. Math.}\ }\textbf {\bibinfo {volume} {20}},\ \bibinfo
  {pages} {251} (\bibinfo {year} {2012})}\BibitemShut {NoStop}%
\bibitem [{\citenamefont {Silaev}\ and\ \citenamefont
  {Babaev}(2012)}]{silaev2012a}%
  \BibitemOpen
  \bibfield  {author} {\bibinfo {author} {\bibfnamefont {M.}~\bibnamefont
  {Silaev}}\ and\ \bibinfo {author} {\bibfnamefont {E.}~\bibnamefont
  {Babaev}},\ }\href {https://link.aps.org/doi/10.1103/PhysRevB.85.134514}
  {\bibfield  {journal} {\bibinfo  {journal} {Phys. Rev. B}\ }\textbf {\bibinfo
  {volume} {85}},\ \bibinfo {pages} {134514} (\bibinfo {year}
  {2012})}\BibitemShut {NoStop}%
\bibitem [{\citenamefont {Yuan}\ \emph {et~al.}(2009)\citenamefont {Yuan},
  \citenamefont {Singleton}, \citenamefont {Balakirev}, \citenamefont {Baily},
  \citenamefont {Chen}, \citenamefont {Luo},\ and\ \citenamefont
  {Wang}}]{Exp-1}%
  \BibitemOpen
  \bibfield  {author} {\bibinfo {author} {\bibfnamefont {H.~Q.}\ \bibnamefont
  {Yuan}}, \bibinfo {author} {\bibfnamefont {J.}~\bibnamefont {Singleton}},
  \bibinfo {author} {\bibfnamefont {F.~F.}\ \bibnamefont {Balakirev}}, \bibinfo
  {author} {\bibfnamefont {S.~A.}\ \bibnamefont {Baily}}, \bibinfo {author}
  {\bibfnamefont {G.~F.}\ \bibnamefont {Chen}}, \bibinfo {author}
  {\bibfnamefont {J.~L.}\ \bibnamefont {Luo}}, \ and\ \bibinfo {author}
  {\bibfnamefont {N.~L.}\ \bibnamefont {Wang}},\ }\href@noop {} {\bibfield
  {journal} {\bibinfo  {journal} {Nature}\ }\textbf {\bibinfo {volume} {457}},\
  \bibinfo {pages} {565 EP } (\bibinfo {year} {2009})}\BibitemShut {NoStop}%
\bibitem [{\citenamefont {Tafti}\ \emph {et~al.}(2014)\citenamefont {Tafti},
  \citenamefont {Clancy}, \citenamefont {Lapointe-Major}, \citenamefont
  {Collignon}, \citenamefont {Faucher}, \citenamefont {Sears}, \citenamefont
  {Juneau-Fecteau}, \citenamefont {Doiron-Leyraud}, \citenamefont {Wang},
  \citenamefont {Luo}, \citenamefont {Chen}, \citenamefont {Desgreniers},
  \citenamefont {Kim},\ and\ \citenamefont {Taillefer}}]{Exp-2}%
  \BibitemOpen
  \bibfield  {author} {\bibinfo {author} {\bibfnamefont {F.~F.}\ \bibnamefont
  {Tafti}}, \bibinfo {author} {\bibfnamefont {J.~P.}\ \bibnamefont {Clancy}},
  \bibinfo {author} {\bibfnamefont {M.}~\bibnamefont {Lapointe-Major}},
  \bibinfo {author} {\bibfnamefont {C.}~\bibnamefont {Collignon}}, \bibinfo
  {author} {\bibfnamefont {S.}~\bibnamefont {Faucher}}, \bibinfo {author}
  {\bibfnamefont {J.~A.}\ \bibnamefont {Sears}}, \bibinfo {author}
  {\bibfnamefont {A.}~\bibnamefont {Juneau-Fecteau}}, \bibinfo {author}
  {\bibfnamefont {N.}~\bibnamefont {Doiron-Leyraud}}, \bibinfo {author}
  {\bibfnamefont {A.~F.}\ \bibnamefont {Wang}}, \bibinfo {author}
  {\bibfnamefont {X.-G.}\ \bibnamefont {Luo}}, \bibinfo {author} {\bibfnamefont
  {X.~H.}\ \bibnamefont {Chen}}, \bibinfo {author} {\bibfnamefont
  {S.}~\bibnamefont {Desgreniers}}, \bibinfo {author} {\bibfnamefont {Y.-J.}\
  \bibnamefont {Kim}}, \ and\ \bibinfo {author} {\bibfnamefont
  {L.}~\bibnamefont {Taillefer}},\ }\href {\doibase 10.1103/PhysRevB.89.134502}
  {\bibfield  {journal} {\bibinfo  {journal} {Phys. Rev. B}\ }\textbf {\bibinfo
  {volume} {89}},\ \bibinfo {pages} {134502} (\bibinfo {year}
  {2014})}\BibitemShut {NoStop}%
\end{thebibliography}%


%

\end{document}